\documentclass[10pt,twocolumn,reprint]{revtex4-1}
\usepackage[T1]{fontenc}
\usepackage[latin9]{inputenc}
\usepackage{amsmath}
\usepackage{amssymb}
\usepackage{graphicx}

\makeatletter

\providecommand{\tabularnewline}{\\}

\usepackage{algcompatible}
\usepackage{placeins}

\makeatother

\begin{document}

\title{Heuristic construction of codeword stabilized codes}

\author{Alex Rigby}
\email{alex.rigby@utas.edu.au}

\author{JC Olivier}

\author{Peter Jarvis}

\affiliation{College of Sciences and Engineering, University of Tasmania, Hobart,
Tasmania 7005, Australia }
\begin{abstract}
The family of codeword stabilized codes encompasses the stabilizer
codes as well as many of the best known nonadditive codes. However,
constructing optimal $n$-qubit codeword stabilized codes is made
difficult by two main factors. The first of these is the exponential
growth with $n$ of the number of graphs on which a code can be based.
The second is the NP-hardness of the maximum clique search required
to construct a code from a given graph. We address the second of these
issues through the use of a heuristic clique finding algorithm. This
approach has allowed us to find $((9,97\leq K\leq100,2))$ and $((11,387\leq K\leq416,2))$
codes, which are larger than any previously known codes. To address
the exponential growth of the search space, we demonstrate that graphs
that give large codes typically yield clique graphs with a large number
of nodes. The number of such nodes can be determined relatively efficiently,
and we demonstrate that $n$-node graphs yielding large clique graphs
can be found using a genetic algorithm. This algorithm uses a novel
spectral bisection based crossover operation that we demonstrate to
be superior to more standard crossover operations. Using this genetic
algorithm approach, we have found $((13,18,4))$ and $((13,20,4))$
codes that are larger than any previously known code. We also consider
codes for the amplitude damping channel. We demonstrate that for $n\leq9$,
optimal codeword stabilized codes correcting a single amplitude damping
error can be found by considering standard form codes that detect
one of only three of the $3^{n}$ possible equivalent error sets.
By combining this error set selection with the genetic algorithm approach,
we have found $((11,68))$ and $((11,80))$ codes capable of correcting
a single amplitude damping error and $((11,4))$, $((12,4))$, $((13,8))$,
and $((14,16))$ codes capable of correcting two amplitude damping
errors.
\end{abstract}
\maketitle

\section{Introduction }

Quantum codes can be used to protect quantum information against the
effects of a noisy channel. An $n$-qubit code is a subspace $\mathcal{Q}\subseteq(\mathbb{C}^{2})^{\otimes n}$
of dimension $K$. If $\mathcal{Q}$ can detect any arbitrary error
on fewer than $d$ qubits, but not some error on $d$ qubits, then
$\mathcal{Q}$ is said to have distance $d$ and is called an $((n,K))$
or $((n,K,d))$ code. Equivalently, a code has distance $d$ if it
can detect the set $\mathcal{E}$ of all Pauli errors of weight less
than $d$ but cannot detect some weight $d$ Pauli error \citep{knill1997theory}.
A well understood family of codes is the stabilizer (additive) codes,
which are codes defined using an abelian subgroup of the $n$-qubit
Pauli group \citep{gottesman1997stabilizer}. However, codes outside
of the stabilizer framework, called nonadditive codes, can potentially
encode a larger subspace while still detecting the same error set
\citep{rains1997nonadditive,yu2008nonadditive,yu2007graphical,rains1999quantum,smolin2007simple,grassl2008quantum}.
Codeword stabilized (CWS) codes encompass both the stabilizer codes
and many of the best known nonadditive codes \citep{cross2008codeword,chuang2009codeword}.
In general, an $((n,K))$ CWS code is defined using an $n$-qubit
stabilizer state, which is a stabilizer code of dimension one, and
a set of $K$ $n$-qubit Pauli operators called word operators \citep{cross2008codeword}.
A standard form CWS code $\mathcal{Q}$ is one where the stabilizer
state is defined by a simple undirected $n$-node graph $G$ (that
is, it is a graph state) and the word operators are defined by a binary
classical code $\mathcal{C}\subseteq\mathrm{GF}(2)^{n}$ with $|\mathcal{C}|=K$
\citep{cross2008codeword}. For $\mathcal{Q}$ to detect an error
set $\mathcal{E}$, $\mathcal{C}$ must detect the classical error
set $Cl_{G}(\mathcal{E})\subseteq\mathrm{GF}(2)^{n}$ induced by the
graph. An appropriate classical code of maximum size can be found
by constructing a clique graph and performing a maximum clique search
\citep{chuang2009codeword}.

The error set that must be detected by an $((n,K,d))$ code $\mathcal{Q}$
is invariant under any permutation of the Pauli matrices $X$, $Y$,
and $Z$ on any subset of qubits. As a result of this symmetry, we
call $((n,K,d))$ codes symmetric codes. This symmetry also means
that if $\mathcal{Q}'$ is local Clifford (LC) equivalent to $\mathcal{Q}$
(that is, if $\mathcal{Q}'=U\mathcal{Q}$ for some LC operator $U$),
then $\mathcal{Q}'$ is also an $((n,K,d))$ code. It follows from
the LC equivalence of every stabilizer state to a graph state \citep{van2004graphical,grassl2002graphs,schlingemann2001stabilizer}
that every CWS code is LC equivalent to one in standard form \citep{cross2008codeword}.
It is therefore sufficient to consider only standard form codes when
attempting to construct an optimal $((n,K,d))$ CWS code. In fact,
it is sufficient to consider only codes based on graph states that
are not LC equivalent up to a permutation of qubit labels \citep{chuang2009codeword}.
This corresponds to considering only graphs that are not isomorphic
up to a series of local complementations \citep{van2004graphical}.
For $n\leq12$, this set of inequivalent graphs, denoted $\mathcal{L}_{n}$,
has been enumerated \citep{danielsen2005self,danielsen2006classification,danielsen2004database}
and in theory can be exhaustively searched to construct an optimal
code. Such a search of $\mathcal{L}_{10}$ has previously yielded
the well known $((10,24,3))$ code \citep{yu2007graphical}. For distance
two codes, searching $\mathcal{L}_{n}$ quickly becomes prohibitive
with increasing $n$ due to the rapidly growing clique graphs and
the NP-hardness of finding a maximum clique \citep{karp1972reducibility}.
To address this, we employ the heuristic Phased Local Search (PLS)
clique finding algorithm \citep{pullan2006phased}. Using this approach,
we have found $((9,97\leq K\leq100,2))$ and $((11,387\leq K\leq416,2))$
codes that are larger than the best known nonadditive codes presented
in Ref. \citep{rains1999quantum} and Ref. \citep{smolin2007simple}
respectively.

The apparent exponential growth of $|\mathcal{L}_{n}|$ with increasing
$n$ means that even if $\mathcal{L}_{n}$ were enumerated for $n\geq13$,
an exhaustive search would be prohibitive. As such, other search strategies
are required for constructing codes. To aid this search, we demonstrate
a relationship between the code size and the order (number of nodes)
of the clique graph yielded by a given graph. In particular, we show
that the clique graph orders exhibit clustering and that the graphs
yielding the best codes tend to belong to the highest clique graph
order cluster. This reduces the search to finding graphs that yield
large clique graphs, and we show that such graphs can be generated
using a genetic algorithm to search the set of all distinct $n$-node
graphs. This genetic algorithm uses a novel crossover operation based
on spectral bisection, which we show to be significantly more effective
than standard single-point, two-point, and uniform crossover operations.
Using this genetic algorithm approach, we have found $((13,18,4))$
and $((13,20,4))$ codes. These codes are larger than an optimal $((13,16,4))$
stabilizer code, and to the best of our knowledge they are the first
$d\geq4$ codes to achieve this (we note that there is a family of
$d=8$ nonadditive codes that are larger than the best known, but
not necessarily optimal, stabilizer codes \citep{grassl2008quantum}).

For asymmetric codes, the error set $\mathcal{E}$ that they detect
is no longer invariant under Pauli matrix permutation. This means
that if $\mathcal{Q}$ detects $\mathcal{E}$, then there is no guarantee
that an LC-equivalent code $\mathcal{Q}'=U\mathcal{Q}$ also detects
$\mathcal{E}$. However, if $\mathcal{Q}$ detects the LC-equivalent
error set $U^{\dagger}\mathcal{E}U$, then $\mathcal{Q}'$ will detect
$\mathcal{E}$. As a result, when attempting to construct an optimal
$((n,K))$ code CWS code detecting $\mathcal{E}$, it is sufficient
to consider standard form codes based on elements of $\mathcal{L}_{n}$
that detect one of the up to $6^{n}$ possible LC-equivalent error
sets \citep{jackson2016codeword} (the $6^{n}$ value stems from there
being six possible permutations of the Pauli matrices on each of the
$n$ qubits). Such an asymmetric error set arises when constructing
codes that correct amplitude damping errors. In this case, a partial
symmetry reduces the number of LC-equivalent error sets to $3^{n}$;
however, this is still large enough to make an exhaustive search prohibitive
for $n\geq10$. Again, we therefore require different search strategies
for constructing codes. We demonstrate that for $n\leq9$, optimal
CWS codes correcting a single amplitude damping error can be found
by considering only codes based on nonisomorphic graphs that detect
one of three LC-equivalent error sets. By combining this error set
selection with the genetic algorithm approach, we have found $((11,68))$
and $((11,80))$ codes capable of correcting a single amplitude damping
error. These are larger than the best known stabilizer codes detecting
the same error set \citep{gottesman1997stabilizer}. We have also
found $((11,4))$, $((12,4))$, $((13,8))$, and $((14,16))$ stabilizer
codes capable of correcting two amplitude damping errors.

The paper is organized as follows. Section \ref{sec:Background} gives
an introduction to undirected graphs, genetic algorithms, classical
codes, and quantum codes. Section \ref{sec:Symmetric-codes} details
our search strategies for symmetric codes and presents the new codes
we have found. This is then extended to asymmetric codes for the amplitude
damping channel in Sec. \ref{sec:Asymmetric-codes}. The paper is
concluded in Sec. \ref{sec:Conclusion}.

\section{Background\label{sec:Background}}

\subsection{Undirected graphs\label{subsec:Undirected-graphs}}

\begin{figure}
\includegraphics[scale=0.55]{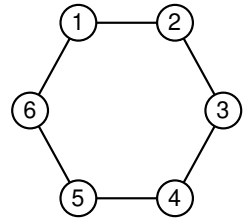}\caption{\label{fig:cycle graph depiction}A drawing of a cycle graph where
the circles correspond to nodes and the lines to edges.}
\end{figure}

A simple undirected graph $G=(N,E)$ of order $n$ consists of a set
of nodes $N=\{v_{1},v_{2},\dots,v_{n}\}$ and a set of edges $E\subseteq\{\{v_{i},v_{j}\}:v_{i},v_{j}\in N,v_{i}\neq v_{j}\}$.
An edge $e=\{v_{i},v_{j}\}\in E$ is an unordered pair that connects
the nodes $v_{i},v_{j}\in N$, which are called the endpoints of $e$.
A graph is typically drawn with the nodes depicted as circles that
are joined by lines representing the edges. An example of such a drawing
is given in Fig. \ref{fig:cycle graph depiction}. $G$ can be represented
by the symmetric $n\times n$ adjacency matrix $\Gamma$ where
\begin{equation}
\Gamma_{ij}=\begin{cases}
1 & \mathrm{if}\,\{v_{i},v_{j}\}\in E,\\
0 & \mathrm{otherwise}.
\end{cases}
\end{equation}
The neighborhood $\mathcal{N}(v_{i})=\{v_{j}:\{v_{i},v_{j}\}\in E\}$
of some node $v_{i}\in N$ is the set of nodes to which it is connected.
The degree $\deg(v_{i})=|\mathcal{N}(v_{i})|$ of $v_{i}$ is the
number of nodes to which it is connected. The $n\times n$ degree
matrix $D$ has elements
\begin{equation}
D_{ij}=\begin{cases}
\deg(v_{i}) & \mathrm{if}\,i=j,\\
0 & \mathrm{otherwise}.
\end{cases}
\end{equation}

A subgraph $G_{S}(N_{S},E_{S}$) of $G=(N,E)$ is a graph with nodes
$N_{S}\subseteq N$ and edges $E_{S}\subseteq E$. The subgraph induced
by a subset of nodes $N_{I}\subseteq N$ is the graph $G_{I}=G[N_{I}]=(N_{I},E_{I})$
where $E_{I}=\{\{v_{i},v_{j}\}\in E:v_{i},v_{j}\in N_{I}\}$ contains
all the edges in $E$ that have both endpoints in $N_{I}$. A walk
is a sequence whose elements alternate between connected nodes and
the edges that connect them. For example, $1,\{1,2\},2,\{2,3\},3,\{3,4\},4,\{3,4\},3$
is a walk in the graph shown in Fig. \ref{fig:cycle graph depiction}.
The length of a walk is the number of edges it contains. A path is
a walk containing no repeated nodes or edges with the exception that
the first and last node can be the same, in which case the path is
called a cycle. A graph such as the one shown in Fig. \ref{fig:cycle graph depiction}
where all nodes belong to a single cycle is called a cycle graph.
A graph is connected if there is a path between any two of its nodes.
A connected component of $G$ is a maximal connected subgraph $G_{S}(N_{S},E_{S})$
{[}maximal in that there is no other connected subgraph $G_{T}(N_{T},E_{T})$
where $N_{S}\subset N_{T}${]}.

Two graphs $G_{1}=(N_{1},E_{1})$ and $G_{2}=(N_{2},E_{2})$ are isomorphic
if they differ only up to a relabeling of nodes. Formally, they are
isomorphic if there exists an isomorphism from $G_{1}$ to $G_{2}$,
which is a bijection $f:N_{1}\rightarrow N_{2}$ such that $\{v_{i},v_{j}\}\in E_{1}$
if and only if $\{f(v_{i}),f(v_{j})\}\in E_{2}$. An isomorphism $f:N\rightarrow N$
from a graph $G=(N,E)$ to itself is called an automorphism. The set
of all all automorphisms of $G$ forms a group $\mathrm{Aut}(G)$
under composition. There are a number of packages, such as NAUTY \citep{mckay1981practical,mckay2013nauty},
available for determining the automorphism group of a given graph.
We denote the set of all distinct $n$-node graphs with nodes $N=\{1,2,\dots,n\}$
as $\mathcal{D}_{n}$, the size of which grows exponentially with
$|\mathcal{D}_{n}|=2^{\binom{n}{2}}$. $\mathcal{D}_{n}$ can be partitioned
up to isomorphism to give the set $\mathcal{G}_{n}$; $|\mathcal{G}_{n}|$
also grows exponentially with $n$ \citep{harary2014graphical}, as
shown in Table \ref{graph number table} for $n\leq12$. The size
of some $g\in\mathcal{G}_{n}$ with representative $G\in\mathcal{D}_{n}$
is $n!/|\mathrm{Aut}(G)|$ \citep{harary2014graphical}. 

\begin{table}
\caption{\label{graph number table}The number of distinct, nonisomorphic,
and non-LC-isomorphic graphs with $n\protect\leq12$ nodes.}
\begin{tabular}{c|ccc}
$n$ & $|\mathcal{D}_{n}|$ & $|\mathcal{G}_{n}|$ & $|\mathcal{L}_{n}|$\tabularnewline
\hline 
$1$ & $2^{0}$ & $1$ & $1$\tabularnewline
$2$ & $2^{1}$ & $2$ & $2$\tabularnewline
$3$ & $2^{3}$ & $4$ & $3$\tabularnewline
$4$ & $2^{6}$ & $11$ & $6$\tabularnewline
$5$ & $2^{10}$ & $34$ & $11$\tabularnewline
$6$ & $2^{15}$ & $156$ & $26$\tabularnewline
$7$ & $2^{21}$ & $1,044$ & $59$\tabularnewline
$8$ & $2^{28}$ & $12,346$ & $182$\tabularnewline
$9$ & $2^{36}$ & $274,668$ & $675$\tabularnewline
$10$ & $2^{45}$ & $12,005,168$ & $3,990$\tabularnewline
$11$ & $2^{55}$ & $1,018,997,864$ & $45,144$\tabularnewline
$12$ & $2^{66}$ & $165,091,172,592$ & $1,323,363$\tabularnewline
\end{tabular}
\end{table}

The complement $\bar{G}=(N,\bar{E})$ of a graph $G=(N,E)$ has an
edge $\{v_{i},v_{j}\}\in\bar{E}$ if and only if $\{v_{i},v_{j}\}\notin E$.
A local complementation (LC) at node $v_{i}$ replaces the induced
subgraph $G[\mathcal{N}(v_{i})]$ with its complement. If two graphs
$G_{1},G_{2}\in\mathcal{D}_{n}$ differ by a series of local complementations,
then we say they are LC equivalent. If a series of local complementations
applied to $G_{1}$ yields a graph $G_{2}'$ that is isomorphic to
$G_{2}$, then we say that $G_{1}$ and $G_{2}$ are LC-isomorphic.
Partitioning $\mathcal{D}_{n}$ up to LC-isomorphism gives the set
$\mathcal{L}_{n}$, which has been enumerated for $n\leq12$ \citep{danielsen2005self,danielsen2006classification,danielsen2004database}
and also seems to grow exponentially with $n$ as shown in Table \ref{graph number table}.
Any two graphs that are isomorphic are necessarily LC-isomorphic,
and therefore any element $l\in\mathcal{L}_{n}$ is the union $l=\cup_{i}g_{i}$
of elements $g_{i}\in\mathcal{G}_{n}$. These $g_{i}$ can be determined
from any representative of $l$ using Algorithm 5.1 of Ref. \citep{danielsen2005self}.
If a subset $A\subseteq\mathcal{D}_{n}$ contains graphs that are
representatives of $m$ different elements of $\mathcal{G}_{n}$ ($\mathcal{L}_{n}$),
then we say $m$ of the graphs in $A$ are nonisomorphic (non-LC-isomorphic). 

A graph $G=(N,E)$ is complete if every node is connected to every
other node; that is, if $E=\{\{v_{i},v_{j}\}:v_{i},v_{j}\in N,v_{i}\neq v_{j}\}$.
If an induced subgraph $G[\tilde{N}]$ for some $\tilde{N}\subseteq N$
is complete, then $\tilde{N}$ is called a clique. A clique of maximum
size in $G$ is called a maximum clique. Finding a maximum clique
in a graph is is NP-hard \citep{karp1972reducibility}; however, there
are a number of heuristic algorithms that can find large, if not maximum,
cliques. One such algorithm is the Phased Local Search (PLS) \citep{pullan2006phased},
which performs well compared to other heuristic algorithms in terms
of both speed and clique finding ability \citep{wu2015review}. The
PLS algorithm constructs a clique by initially selecting a node at
random. It then iteratively selects nodes to add to the current clique
(potentially replacing an existing node in the clique) until a maximum
number of selections is reached. To ensure good performance on graphs
with varying structures, PLS cycles through multiple different selection
methods. The search is repeated for a prescribed number of attempts,
after which the largest clique found is returned.

A bipartition of $G=(N,E)$ divides the nodes into two disjoint subsets
$N_{1}$ and $N_{2}$. A bipartition is called a bisection if $|N_{1}|=|N_{2}|$
for even $|N|$ or if $||N_{1}|-|N_{2}||=1$ for odd $|N|$. An optimal
bisection is one that minimizes the number of edges connecting nodes
in $N_{1}$ to those in $N_{2}$. Finding such an optimal bisection
is NP-hard \citep{garey2002computers}; however, approximate heuristic
approaches are available. One such approach is spectral bisection
\citep{hall1970r,donath1972algorithms,fiedler1975property,pothen1990partitioning},
which is based on the graph's Laplacian matrix $L=D-\Gamma$. $L$
is positive semidefinite and as such has real, nonnegative eigenvalues.
The eigenvector $\boldsymbol{u}=(u_{1},\dots,u_{n})$ corresponding
to the second smallest eigenvalue is called the Fiedler vector \citep{fiedler1973algebraic}.\textbf{
}The Fiedler vector can be used to bisect $N$, with the indices of
the $\lfloor n/2\rfloor$ smallest components of $\boldsymbol{u}$
dictating the nodes in $N_{1}$ and $N_{2}$ simply being $N_{2}=N\backslash N_{1}$.

\subsection{Genetic algorithms\label{subsec:Genetic-algorithms}}

Suppose we wish to determine which element in a set $A$ is optimal
in some sense. This can be expressed as finding the $a\in A$ that
maximizes a fitness function $f:A\rightarrow\mathbb{R}$. The brute
force approach to this problem is to determine the fitness of every
element $a\in A$. This is called an exhaustive search and quickly
becomes impractical if the search space $A$ is large and/or evaluating
the fitness of elements is computationally intensive. In such cases,
heuristic search algorithms can be used to find good, but potentially
not optimal, elements of $A$. The simplest such approach is a random
search, where fitness is calculated only for the elements in a randomly
selected subset $B\subset A$. Another heuristic search strategy is
the genetic algorithm, which is inspired by natural evolution \citep{whitley1994genetic,lones2011sean}.
There are many genetic algorithm variants; a simple implementation
is as follows. Initially, the child population, which is an $N$ element
subset of $A$, is randomly generated. This is followed by a calculation
of each child's fitness (a child being an element of the child population).
The genetic algorithm then iterates for some predetermined number
of maximum generations. In each generation the previous generation's
child population becomes the current generation's parent population
(whose elements are called parents). A new child population is then
formed by selecting two parents at a time and producing two children
from them. The parents are selected according to their fitness, with
high fitness parents having a higher chance of selection. With probability
$p_{c}$, the two children will be produced via crossover, which combines
attributes of the two parents; otherwise, they will simply be duplicates
of their parents. Each child is then subjected to mutation (random
alteration) with probability $p_{m}$ before being added to the child
population. Once the child population again contains $N$ children,
their fitnesses are calculated and a new generation begins.

Tournament selection is a simple and commonly used method of selecting
parents based on their fitness. First, a subset of the parent population
is chosen at random, then the fittest parent within this subset is
selected. The size of the subset chosen is called the tournament size;
it controls the selection pressure of the genetic algorithm, which
is a measure of how dependent selection is on having high fitness.
If the tournament size is large, then there is high selection pressure,
meaning that the highest fitness parents tend to be selected. This
greedy/exploitative approach gives faster convergence; however, the
search is more likely to become stuck at a suboptimal local maximum
\citep{legg2004tournament}. Conversely, a small tournament size will
lead to greater exploration of the search space at the cost of potentially
slow convergence. A common modification to the genetic algorithm is
the inclusion of elitist selection, which involves adding some number
of the fittest parents to the child population at the start of each
generation. This preserves the best elements; however, the increased
selection pressure can again increase the probability of convergence
to a suboptimal local maximum.

The crossover and mutation operations used depend on how elements
of $A$ are represented. A standard representation involves encoding
elements as bit strings of fixed length $b$. A common and simple
mutation operation in this case involves flipping any given bit in
a child bit string with some probability (this probability is often
taken to be $1/b$ \citep{lones2011sean}). Standard crossover methods
include single-point, two-point, and uniform crossover. In single-point
crossover, an index $1\leq i\leq b$ is chosen, and the values beyond
this point are exchanged between the two parent bit strings to form
two child bit strings. In two-point crossover, two such indices are
selected and all values between them are exchanged. In uniform crossover,
each individual bit is exchanged between the two parents with some
probability $p_{e}$.

In some cases, representations other than bit strings are more natural.
For example, it may be possible represent elements as graphs. Crossover
becomes more complicated with such a representation. A potential method
is presented in Ref. \citep{globus1999automatic} and is as follows.
First, each of the parent graphs $P_{1}$ and $P_{2}$ are each split
into two subgraphs, called fragments, to produce disconnected parents
$P_{1D}$ and $P_{2D}$. To split a parent graph, first an edge $\{v_{i},v_{j}\}$
is chosen at random. In an iterative process, the shortest path between
$v_{i}$ and $v_{j}$ is determined, and a randomly selected edge
in this path is removed (in the first iteration, this will simply
be the edge $\{v_{i},v_{j}\}$). This continues until no path exists
between $v_{i}$ and $v_{j}$. The connected component containing
$v_{i}$ is the fragment $F_{1}$, and the subgraph induced by the
remaining nodes is the fragment $F_{2}$. In the next step, disconnected
children $C_{1D}$ and $C_{2D}$ are formed by exchanging a fragment,
say $F_{1}$, between of each of the parent graphs. The two fragments
in each disconnected child are then combined to produce children $C_{1}$
and $C_{2}$. This combination process involves iteratively selecting
a node from each fragment and joining them with an edge. The probability
of a node being selected is proportional to the difference in its
current degree to its degree in its initial parent graph. This process
of adding edges is repeated until all of the nodes in one of the fragments,
say $F_{1}$, have the same degree as they did in their initial parent
graph. If a node $v_{l}$ in $F_{2}$ has degree lower than its initial
degree by some amount $\delta_{l}$, then in a process repeated $\delta_{l}$
times, it will be connected to a randomly selected node in $F_{1}$
with $50\%$ probability. As outlined in Ref. \citep{stone2011crossover},
the splitting process presented here has some undesirable attributes.
Firstly, it tends to produce two fragments with a vastly different
number of nodes. Secondly, it often removes a large number of edges
from within the larger fragment; these are edges that did not have
to be removed to split the parent graph.

\subsection{Classical codes\label{subsec:Classical-codes}}

A classical channel is a map $\Phi:\mathcal{A}_{x}\rightarrow\mathcal{A}_{y}$,
where $\mathcal{A}_{x}$ is the set of possible inputs and $\mathcal{A}_{y}$
is the set of possible outputs. We are concerned with channels where
the input and outputs are binary; that is, $\mathcal{A}_{x}=\mathcal{A}_{y}=\mathrm{GF}(2)$.
In this case, the action of the channel can be expressed as
\begin{equation}
\Phi(x)=x+e=y,
\end{equation}
where $x\in\mathrm{GF}(2)$ is the channel input, $y\in\mathrm{GF}(2)$
is the channel output, and $e\in\mathrm{GF}(2)$ is an error (or noise)
symbol. A code can be used to protect against the noise introduced
by the channel. A length $n$ binary code is a subset $\mathcal{C}\subseteq\mathrm{GF}(2)^{n}$
whose elements are called codewords. Codewords are transmitted as
$n$ sequential uses of $\Phi$ or, equivalently, as a single use
of the combined channel $\Phi^{n}$, which is comprised of $n$ copies
of $\Phi$. The action of $\Phi^{n}$ on some input $\boldsymbol{x}\in\mathcal{C}$
is
\begin{equation}
\Phi^{n}(\boldsymbol{x})=\boldsymbol{x}+\boldsymbol{e}=\boldsymbol{y},
\end{equation}
where $\boldsymbol{y}\in\mathrm{GF}(2)^{n}$ is the channel output
and $\boldsymbol{e}\in\mathrm{GF}(2)^{n}$ is an error ``vector''.
The weight of an error is the number of nonzero components from which
it comprised. 

We say that a code $\mathcal{C}$ can detect a set of errors $\mathcal{E}\subseteq\mathrm{GF}(2)$
if
\begin{equation}
\boldsymbol{x}_{i}+\boldsymbol{e}\neq\boldsymbol{x}_{j}\label{eq:classical error correction criteria}
\end{equation}
for all $\boldsymbol{e}\in\mathcal{E}$ and $\boldsymbol{x}_{i},\boldsymbol{x}_{j}\in\mathcal{C}$,
where $\boldsymbol{x}_{i}\neq\boldsymbol{x}_{j}$. That is, the errors
in $\mathcal{E}$ can be detected if they do not map one codeword
to another. Furthermore, we say that $\mathcal{C}$ can correct $\mathcal{E}$
if
\begin{equation}
\boldsymbol{x}_{i}+\boldsymbol{e}_{k}\neq\boldsymbol{x}_{j}+\boldsymbol{e}_{l}\label{eq:classical correction criteria}
\end{equation}
for all $\boldsymbol{e}_{k},\boldsymbol{e}_{l}\in\mathcal{E}$ and
$\boldsymbol{x}_{i},\boldsymbol{x}_{j}\in\mathcal{C}$, where $\boldsymbol{x}_{i}\neq\boldsymbol{x}_{j}$.
This condition simply ensures that two codewords cannot be mapped
to the same $\boldsymbol{y}\in\mathrm{GF}(2)^{n}$, in which case
the transmitted codeword cannot be inferred with certainty. $\mathcal{C}$
is said to have distance $d$ if it can detect any error of weight
less than $d$ but is unable to detect some weight $d$ error. Note
that $\mathcal{C}$ can correct $\mathcal{E}$ if and only if it can
detect $\mathcal{E}+\mathcal{E}=\{\boldsymbol{e}_{k}+\boldsymbol{e}_{l}:\boldsymbol{e}_{k},\boldsymbol{e}_{l}\in\mathcal{E}\}$,
meaning that a distance $d$ code can correct any error of weight
$t=\lfloor(d-1)/2\rfloor$ or less. A length $n$ code $\mathcal{C}$
of size $|\mathcal{C}|=K$ and distance $d$ is called an $(n,K)$
or $(n,K,d)$ code. If $\mathcal{C}$ forms a vector space, then it
is called linear and has $K=2^{k}$. A linear code encodes the state
of $k$ bits and is called an $[n,k]$ or $[n,k,d]$ code.

Finding a code $\mathcal{C}$ of maximum size that detects an error
set $\mathcal{E}$ can be expressed as a clique finding problem. This
is achieved by constructing a graph $G_{\mathcal{E}}=(N_{\mathcal{E}},E_{\mathcal{E}})$
whose nodes are potential codewords; that is, $N_{\mathcal{E}}=\mathrm{GF}(2)^{n}$.
Two nodes $\boldsymbol{x}_{i},\boldsymbol{x}_{j}\in N_{\mathcal{E}}$
are connected by an edge $\{\boldsymbol{x}_{i},\boldsymbol{x}_{j}\}\in E_{\mathcal{E}}$
if $\boldsymbol{x}_{i}+\boldsymbol{x}_{j}\notin\mathcal{E}$ (that
is, if there is not an error mapping one to the other). Any clique
$\mathcal{C}$ in $G_{\mathcal{E}}$ is a code detecting $\mathcal{E}$,
and a code of maximum possible size is a maximum clique in $G_{\mathcal{E}}$.
Note that if a code $\mathcal{C}$ detects $\mathcal{E}$, then so
does $\mathcal{C}'=\boldsymbol{x}+\mathcal{C}$ for any $\boldsymbol{x}\in\mathcal{C}$.
As $\boldsymbol{0}\in\mathcal{C}'$ and $|\mathcal{C}|=|\mathcal{C}'|$,
this means there is always an optimal code (that is, a maximum size
code detecting $\mathcal{E}$) that contains the all zero codeword.
The clique search can be restricted to such codes by taking $N_{\mathcal{E}}=\mathrm{GF}(2)^{n}\backslash\mathcal{E}$.

\subsection{Quantum codes\label{subsec:Quantum-channels-and}}

The action of a quantum channel $\Phi$ on a quantum state described
by the density operator $\rho$ is
\begin{equation}
\Phi(\rho)=\sum_{k}A_{k}\rho A_{k}^{\dagger},
\end{equation}
where the $A_{k}$, called Kraus operators, satisfy $\sum_{k}A_{k}^{\dagger}A_{k}=I$
(the identity operator) \citep{kraus1983states}. The channel can
be interpreted as mapping $\rho\mapsto A_{k}\rho A_{k}^{\dagger}$
(up to normalization) with probability $\mathrm{tr}(A_{k}\rho A_{k}^{\dagger})$
\citep{nielsen2002quantum}. If $\rho=|\phi\rangle\langle\phi|$ (that
is, if the input state is pure), then then this becomes the mapping
$|\phi\rangle\mapsto A_{k}|\phi\rangle$ (up to normalization) with
corresponding probability $\langle\phi|A_{k}^{\dagger}A_{k}|\phi\rangle$.
In this paper we are interested in qubit systems; that is, systems
where states $|\phi\rangle$ belong to a two dimensional Hilbert space
$\mathcal{H}\cong\mathbb{C}^{2}$. Similar to the classical case,
the noise introduced by a quantum channel can be protected against
by employing a code. A quantum (qubit) code of length $n$ is a subspace
$\mathcal{Q}\subseteq(\mathbb{C}^{2})^{\otimes n}$. Codewords $|\phi\rangle\in\mathcal{Q}$
are transmitted across the combined $n$-qubit channel $\Phi^{\otimes n}$.

Suppose a code $\mathcal{Q}$ has an orthonormal basis $\mathcal{B}=\{|\phi_{1}\rangle,\dots,|\phi_{K}\rangle\}$,
and take $\mathcal{E}=\{E_{1},\dots,E_{r}\}$ to be the basis for
some complex vector space of linear $n$-qubit operators (called error
operators). We say that $\mathcal{Q}$ can detect any error in the
span of $\mathcal{E}$ if
\begin{equation}
\langle\phi_{i}|E|\phi_{j}\rangle=C_{E}\delta_{ij}\label{error detection criteria}
\end{equation}
for all $E\in\mathcal{E}$ and $|\phi_{i}\rangle,|\phi_{j}\rangle\in\mathcal{B}$,
where $C_{E}$ is a scalar that depends only on $E$ \citep{gottesman1997stabilizer}.
Furthermore, we say that $\mathcal{Q}$ can correct any error in the
span of $\mathcal{E}$ if 
\begin{equation}
\langle\phi_{i}|E_{k}^{\dagger}E_{l}|\phi_{j}\rangle=C_{kl}\delta_{ij}\label{error correction criteria}
\end{equation}
for all $E_{k},E_{l}\in\mathcal{E}$ and $|\phi_{i}\rangle,|\phi_{j}\rangle\in\mathcal{B}$,
where $C$ is an $r\times r$ Hermitian matrix \citep{knill1997theory}.
The weight of an error $E$ is the number of qubits on which it acts.
$\mathcal{Q}$ has distance $d$ if it can detect any error of weight
less than $d$ but not some weight $d$ error. Similar to the classical
case, a code can correct $\mathcal{E}$ if and only if it can detect
$\mathcal{E}^{\dagger}\mathcal{E}=\{E_{k}^{\dagger}E_{l}:E_{k},E_{l}\in\mathcal{E}\}$,
meaning that a distance $d$ quantum code can also correct any error
of weight $t=\lfloor(d-1)/2\rfloor$ or less. A length $n$ code of
dimension $K$ and distance $d$ is called an $((n,K))$ or $((n,K,d))$
code (the double brackets differentiate from the classical case).
A code $\mathcal{Q}$ correcting $\mathcal{E}$ is called nondegenerate
if the spaces $E_{k}Q$ and $E_{l}Q$ are linearly independent (that
is, their intersection is trivial) for any $E_{k},E_{l}\in\mathcal{E}$,
where $E_{k}\neq E_{l}$. If all such spaces are orthogonal, then
$\mathcal{Q}$ is called pure.

The Pauli matrices in the computational $\{|0\rangle,|1\rangle\}$
basis are
\begin{equation}
X=\left(\begin{array}{cc}
0 & 1\\
1 & 0
\end{array}\right),\,Y=\left(\begin{array}{cc}
0 & -i\\
i & 0
\end{array}\right),\,Z=\left(\begin{array}{cc}
1 & 0\\
0 & -1
\end{array}\right).
\end{equation}
$X$ can be viewed as a bit flip operator as $X|0\rangle=|1\rangle$
and $X|1\rangle=|0\rangle$. $Z$ can be viewed as a phase flip as
$Z|0\rangle=|0\rangle$ and $Z|1\rangle=-|1\rangle$ . $Y=iXZ$ can
be viewed as a combined bit and phase flip. The Pauli matrices are
Hermitian, unitary, and anticommute with each other. Furthermore,
they form a group called the Pauli group
\begin{equation}
\mathcal{P}_{1}=\{\pm I,\pm iI,\pm X,\pm iX,\text{\ensuremath{\pm Y,\pm iY,\pm Z,\pm iZ}\}=\ensuremath{\langle X,Y,Z\rangle}.}
\end{equation}
The $n$-qubit Pauli group $\mathcal{P}_{n}$ is defined as all $n$-fold
tensor product combinations of elements of $\mathcal{P}_{1}$. For
example, $\mathcal{P}_{8}$ contains the element $I\otimes I\otimes X\otimes I\otimes Y\otimes Z\otimes I\otimes I$,
which we can write more compactly as $X_{3}Y_{5}Z_{6}$. The weight
$w(g)$ of some $g\in\mathcal{P}_{n}$ is the number of elements in
the tensor product that are not equal to the identity up to phase.
The commutation relations of the Pauli matrices mean that elements
of $\mathcal{P}_{n}$ must either commute or anticommute, with two
elements anticommuting if their nonidentity components differ in an
odd number of places. The Pauli matrices along with the identity form
a basis for the complex vector space of all $2\times2$ matrices.
It therefore follows that 
\begin{equation}
\mathcal{E}^{r}=\{E=\sigma_{1}\otimes\dots\otimes\sigma_{n}:\sigma_{i}\in\{I,X,Y,Z\}\,\mathrm{and}\,w(E)\leq r\},\label{eq:pauli basis}
\end{equation}
is a basis for all $n$-qubit errors of weight less than or equal
to $r$. An equivalent definition is $\mathcal{E}^{r}=\{E_{1}\dots E_{r}:E_{i}\in\mathcal{E}^{1}\}$;
that is, $\mathcal{E}_{r}$ is the set of all $r$-fold products of
elements of $\mathcal{E}_{1}$, which can be written as $\mathcal{E}^{1}=\{I,X_{i},Y_{i},Z_{i}\}$
where $1\leq i\leq n$. It is sometimes convenient to express some
$E\in\mathcal{P}_{n}$ up to phase as $E\propto X^{u_{1}}Z^{v_{1}}\otimes\dots\otimes X^{u_{n}}Z^{v_{n}}=X^{\boldsymbol{u}}Z^{\boldsymbol{v}}$
where $\boldsymbol{u}=(u_{1},\dots,u_{n}),\boldsymbol{v}=(v_{1},\dots,v_{n})\in\mathrm{GF}(2)^{n}$. 

Two $n$-qubit codes $\mathcal{Q}$ and $\mathcal{Q}'$ are local
unitary (LU) equivalent if $\mathcal{Q}'=U\mathcal{Q}$ for some $U\in U(2)^{\otimes n}$.
These codes will have the same dimension as if $\mathcal{B}=\{|\phi_{1}\rangle,\dots,|\phi_{K}\rangle\}$
is an orthonormal basis for $\mathcal{Q}$, then $\mathcal{B}'=U\mathcal{B}=\{U|\phi_{1}\rangle,\dots,U|\phi_{K}\rangle\}$
is an orthonormal basis for $\mathcal{Q}'$. It follows from Eq. (\ref{error detection criteria})
that $\mathcal{Q}'$ detects the error set $\mathcal{E}$ if and only
if $\mathcal{Q}$ detects the LU-equivalent error set $\mathcal{E}'=U^{\dagger}\mathcal{E}U$.
Furthermore, $\mathcal{E}$ is a basis for all errors of weight less
than $d$ if and only if $\mathcal{E}'$ is also such a basis. Therefore,
$\mathcal{Q}$ and $\mathcal{Q}'$ have the same distance; that is,
they are both $((n,K,d))$ codes. If two codes differ by a LU operator
and/or permutation of qubit labels, which also has no effect on the
size or distance of the code, then they are called equivalent codes.
The normalizer of $\mathcal{P}_{1}$ in $U(2)$ is the single-qubit
Clifford group $\mathcal{C}_{1}=\{U\in U(2):U^{\dagger}\mathcal{P}_{1}U=\mathcal{P}_{1}\}$.
The $n$-qubit local Clifford group $\mathcal{C}_{1}^{n}$ is comprised
of all possible $n$-fold tensor products of elements from $\mathcal{C}_{1}$.
Two codes are local Clifford (LC) equivalent if they are LU equivalent
for some $U\in\mathcal{C}_{1}^{n}\subset U(2)^{\otimes n}$.

Stabilizer codes (also called additive codes) are defined by an abelian
subgroup $\mathcal{S}<\mathcal{P}_{n}$, called the stabilizer, that
does not contain $-I$ \citep{gottesman1997stabilizer}. The code
$\mathcal{Q}$ is the space of states that are fixed by every element
$s_{i}\in\mathcal{S}$; that is, 
\begin{equation}
\mathcal{Q}=\{|\phi\rangle\in(\mathbb{C}^{2})^{\otimes n}:s_{i}|\phi\rangle=|\phi\rangle\,\forall\,s_{i}\in\mathcal{S}\}.
\end{equation}
The requirement that $-I\notin\mathcal{S}$ both means that no $s\in\mathcal{S}$
can have a phase factor of $\pm i$, and that if $s\in\mathcal{S}$,
then $-s\notin\mathcal{S}$. If $\mathcal{S}$ is generated by $M=\{M_{1},\dots,M_{m}\}\subset\mathcal{P}_{n}$,
then it is sufficient (and obviously necessary) for $\mathcal{Q}$
to be stabilized by every $M_{i}$. Assuming that the set of generators
is minimal, it can be shown that $\dim(\mathcal{Q})=2^{n-m}=2^{k}$
\citep{nielsen2002quantum}; that is, $\mathcal{Q}$ encodes the state
of a $k$-qubit system. An $n$-qubit stabilizer code with dimension
$K=2^{k}$ and distance $d$ is called an $[[n,k]]$ or $[[n,k,d]]$
code.

An $n$-qubit stabilizer state $|\mathcal{S}\rangle$ is an $[[n,0,d]]$
code defined by a stabilizer $\mathcal{S}$ with $n$ generators.
The distance of a stabilizer state is defined to be equal to the weight
of the lowest nonzero weight element in $\mathcal{S}$. A graph state
$|G\rangle$ is a stabilizer state defined by a graph $G\in\mathcal{D}_{n}$.
Each node $i$ corresponds to a qubit and is also associated with
a stabilizer generator
\begin{equation}
M_{i}=X_{i}Z_{\mathcal{N}(i)}=X_{i}\prod_{j\in\mathcal{N}(i)}Z_{j}.
\end{equation}
Each graph state $|G\rangle$ defines a basis $\mathcal{B}=\{Z^{\boldsymbol{w}}|G\rangle:\boldsymbol{w}\in\mathrm{GF}(2)^{n}\}$
for $(\mathbb{C}^{2})^{\otimes n}$ \citep{hein2006entanglement}.
An error $E=X_{i}$ maps the graph state $|G\rangle$ to 
\begin{equation}
X_{i}|G\rangle=X_{i}(X_{i}Z_{\mathcal{N}(i)})|G\rangle=Z_{\mathcal{N}(i)}|G\rangle=Z^{\boldsymbol{r}_{i}}|G\rangle,
\end{equation}
where $\boldsymbol{r}_{i}$ is the $i$th row of the adjacency matrix
for $G$. That is, an $X$ error applied at node $i$ is equivalent
to $Z$ errors being applied at its neighbors; this is called the
$X-Z$ rule \citep{looi2008quantum}. It can be shown that every stabilizer
state is LC equivalent to a graph state \citep{van2004graphical,grassl2002graphs,schlingemann2001stabilizer}.
Two graph states $|G_{1}\rangle$ and $|G_{2}\rangle$ are the same
up to a relabeling of qubits if and only if their corresponding graphs
$G_{1}$ and $G_{2}$ are isomorphic. Furthermore, $|G_{1}\rangle$
and $|G_{2}\rangle$ are LC equivalent if and only if $G_{1}$ and
$G_{2}$ are LC equivalent \citep{van2004graphical}. Therefore, $|G_{1}\rangle$
and $|G_{2}\rangle$ are equivalent (as quantum codes) if $G_{1}$
and $G_{2}$ are LC-isomorphic (the converse does not necessarily
hold as two states can be LU equivalent without being LC equivalent
\citep{ji2007lu}).

\subsection{CWS codes}

The family of codeword stabilized (CWS) codes contains all stabilizer
codes as well as many of the best known nonadditive codes \citep{cross2008codeword,chuang2009codeword}.
An $((n,K))$ CWS code $\mathcal{Q}$ is defined using an $n$-qubit
stabilizer state $|\mathcal{S}\rangle$ and a set of $K$ word operators
$\mathcal{W}=\{W_{1},\dots,W_{K}\}\subset\mathcal{P}_{n}$. In particular,
$\mathcal{Q}$ is the span of the basis codewords $|W_{i}\rangle=W_{i}|\mathcal{S}\rangle$.
Note that for the $|W_{i}\rangle$ to actually form a basis, no two
word operators differ only by a stabilizer element; that is, it cannot
be the case that $W_{i}W_{j}\in\bar{\mathcal{S}}=\cup_{\alpha\in\{\pm1,\pm i\}}\alpha\mathcal{S}$.
For a CWS code, the criterion for detecting an error set $\mathcal{E}$
becomes
\begin{equation}
\langle W_{i}|E|W_{j}\rangle=\langle\mathcal{S}|W_{i}^{\dagger}EW_{j}|\mathcal{S}\rangle=C_{E}\delta_{ij}\label{eq:cws detection criteria}
\end{equation}
for all $E\in\mathcal{E}$ and $W_{i},W_{j}\in\mathcal{W}$. If $\mathcal{E}$
contains only Pauli errors $E\in\mathcal{P}_{n}$, then
\begin{equation}
\langle\mathcal{S}|W_{i}^{\dagger}EW_{j}|\mathcal{S}\rangle=\begin{cases}
0 & \mathrm{if\,}W_{i}^{\dagger}EW_{j}\notin\bar{\mathcal{S}},\\
\alpha & \mathrm{if\,}W_{i}^{\dagger}EW_{j}\in\alpha\mathcal{S},
\end{cases}
\end{equation}
where $\alpha\in\{\pm1,\pm i\}$. Therefore, the $i\neq j$ case of
Eq. (\ref{eq:cws detection criteria}) holds for some $E\in\mathcal{E}$
if and only if
\begin{equation}
W_{i}^{\dagger}EW_{j}\notin\bar{\mathcal{S}}\label{eq:cws non diag condition}
\end{equation}
for all $W_{i},W_{j}\in\mathcal{W}$, where $W_{i}\neq W_{j}$. Furthermore,
the $i=j$ case holds for some $E\in\mathcal{E}$ if and only if either
\begin{equation}
W_{i}^{\dagger}EW_{i}\notin\bar{\mathcal{S}}\label{eq:cws diag condition 1}
\end{equation}
or
\begin{equation}
W_{i}^{\dagger}EW_{i}\in\alpha\mathcal{S}\label{eq:cws diag condition 2}
\end{equation}
for all $W_{i}\in\mathcal{W}$ and some particular $\alpha\in\{\pm1,\pm i\}$.

It follows from the LC equivalence of every stabilizer to a graph
state that every CWS code is LC equivalent to one based on a graph
state $|G\rangle$ with word operators of the form $W_{i}=Z^{\boldsymbol{x}_{i}}$
\citep{cross2008codeword}. Such a code is called a standard form
CWS code, and its basis codewords are simply elements of the graph
basis defined by $G$. The set $\{\boldsymbol{x}_{1},\dots,\boldsymbol{x}_{K}\}\subseteq\mathrm{GF}(2)^{n}$
forms a classical binary code $\mathcal{C}$, and without loss of
generality we can take $\boldsymbol{x}_{1}=\boldsymbol{0}$ \citep{cross2008codeword}.
It can be shown that if $\mathcal{C}$ is linear, then the CWS code
is additive \citep{cross2008codeword}; whereas if $\mathcal{C}$
is not linear, then the code may be additive or nonadditive \citep{chuang2009codeword}
(although if $K\neq2^{k}$, then the CWS code must obviously be nonadditive).
The effect of an error $E\propto X^{\boldsymbol{u}}Z^{\boldsymbol{v}}$
on one of the basis codewords $|W_{i}\rangle=Z^{\boldsymbol{x}_{i}}|G\rangle$
follows from the $X-Z$ rule with
\begin{align}
E|W_{i}\rangle & \propto X^{\boldsymbol{u}}Z^{\boldsymbol{v}}Z^{\boldsymbol{x}_{i}}|G\rangle\nonumber \\
 & \propto Z^{\boldsymbol{v}}Z^{\boldsymbol{x}_{i}}X^{\boldsymbol{u}}|G\rangle\nonumber \\
 & =Z^{\boldsymbol{v}}Z^{\boldsymbol{x}_{i}}Z^{\boldsymbol{u}\Gamma}|G\rangle\nonumber \\
 & =Z^{\boldsymbol{v}}Z^{\boldsymbol{u}\Gamma}Z^{\boldsymbol{x}_{i}}|G\rangle\nonumber \\
 & =Z^{Cl_{G}(E)}|W_{i}\rangle,
\end{align}
where $\Gamma$ is the adjacency matrix for $G$ and
\begin{equation}
Cl_{G}(E\propto X^{\boldsymbol{u}}Z^{\boldsymbol{v}})=\boldsymbol{v}+\boldsymbol{u}\Gamma.\label{eq:pauli to classical map}
\end{equation}
Therefore, the effect of $E\propto X^{\boldsymbol{u}}Z^{\boldsymbol{v}}$
is equivalent to that of $E'=Z^{Cl_{G}(E)}$, where $Cl_{G}(E)\in\mathrm{GF}(2)^{n}$
is a classical error induced by the graph. It follows from this equivalence
that $\langle W_{i}|E|W_{j}\rangle\propto\langle W_{i}|Z^{Cl_{G}(E)}|W_{j}\rangle$,
which means that Eq. (\ref{eq:cws non diag condition}) is satisfied
when
\begin{equation}
Z^{\boldsymbol{x}_{i}}Z^{Cl_{G}(E)}Z^{\boldsymbol{x}_{j}}\notin\bar{\mathcal{S}}.\label{eq:standard form non diag condition}
\end{equation}
For a graph state, the only stabilizer element with no $X$ component
is the identity $I=Z^{\boldsymbol{0}}$. Equation (\ref{eq:standard form non diag condition})
therefore reduces to $\boldsymbol{x}_{i}+Cl_{G}(E)\neq\boldsymbol{x}_{j}$,
which is simply the classical error detection criterion of Eq. (\ref{eq:classical error correction criteria}).
This means that an error $E$ can be detected only if $\mathcal{C}$
detects the classical error $Cl_{G}(E)$. Following the same reasoning,
Eq. (\ref{eq:cws diag condition 1}) becomes $\boldsymbol{x}_{i}+Cl_{G}(E)\neq\boldsymbol{x}_{i}$,
which reduces to $Cl_{G}(E)\neq\boldsymbol{0}$. Equation (\ref{eq:cws diag condition 2})
becomes
\begin{equation}
Z^{\boldsymbol{x}_{i}}EZ^{\boldsymbol{x}_{i}}\in\alpha\mathcal{S},\label{eq:standard form diag}
\end{equation}
which reduces to $E\in\alpha\mathcal{S}$ for $\boldsymbol{x}_{1}=\boldsymbol{0}$.
If there is some $W_{i}=Z^{\boldsymbol{x}_{i}}$ that anticommutes
with $E$, then Eq. (\ref{eq:standard form diag}) becomes $E\in-\alpha\mathcal{S}$.
This would mean that both $\alpha^{-1}E\in\mathcal{S}$ and $-\alpha^{-1}E\in\mathcal{S}$,
from which it follows that $-I\in\mathcal{S}$. This cannot be the
case as $\mathcal{S}$ is a stabilizer. Therefore, to satisfy Eq.
(\ref{eq:cws diag condition 2}) it must be the case that $[Z^{\boldsymbol{x}_{i}},E]=0$
for all $\boldsymbol{x}_{i}\in\mathcal{C}$. For $E\propto X^{\boldsymbol{u}}Z^{\boldsymbol{v}}$,
this condition is equivalent to requiring $\boldsymbol{x}_{i}\cdot\boldsymbol{u}=0$
for all $\boldsymbol{x}_{i}\in\mathcal{C}$, were $\boldsymbol{a}\cdot\boldsymbol{b}=\sum_{i}a_{i}b_{i}$
is the standard Euclidean inner product. In summary, a standard form
code detects $E\propto X^{\boldsymbol{u}}Z^{\boldsymbol{v}}\in\mathcal{E}$
if
\begin{equation}
\boldsymbol{x}_{i}+Cl_{G}(E)\neq\boldsymbol{x}_{j}\label{eq:cws classical condition 1}
\end{equation}
for all $\boldsymbol{x}_{i},\boldsymbol{x}_{j}\in\mathcal{C}$, where
$\boldsymbol{x}_{i}\neq\boldsymbol{x}_{j}$, and either
\begin{align}
Cl_{G}(E) & \neq\boldsymbol{0}\label{eq:cws classical condition 2}
\end{align}
or
\begin{equation}
\boldsymbol{x}_{i}\cdot\boldsymbol{u}=0\label{eq:cws classical condition 3}
\end{equation}
for all $\boldsymbol{x}_{i}\in\mathcal{C}$.

Designing a CWS code $\mathcal{Q}$ for a given graph $G$ and error
set $\mathcal{E}$ consists of finding a classical code $\mathcal{C}$
that satisfies Eqs. (\ref{eq:cws classical condition 1}) to (\ref{eq:cws classical condition 3})
for every $E\in\mathcal{E}$. It is convenient to express this as
a clique finding problem as outlined in Ref. \citep{chuang2009codeword}.
First, the classical set
\begin{equation}
Cl_{G}(\mathcal{E})=\{Cl_{G}(E):E\in\mathcal{E}\}
\end{equation}
induced by the graph is determined. Also required is the set
\begin{align}
D_{G}(\mathcal{E}) & =\{\boldsymbol{x}\in\mathrm{GF}(2)^{n}:Cl_{G}(E)=0\,\text{and}\,\nonumber \\
 & \boldsymbol{x}\cdot\boldsymbol{u}\neq0\,\text{for some}\,E\propto X^{\boldsymbol{u}}Z^{\boldsymbol{v}}\in\mathcal{E}\}.\label{eq:degenerate set}
\end{align}
These are elements of $\mathrm{GF}(2)^{n}$ that cannot be included
in the code as they violate Eqs. (\ref{eq:cws classical condition 2})
and (\ref{eq:cws classical condition 3}). An algorithm for efficiently
determining $Cl_{G}(\mathcal{E})$ and $D_{G}(\mathcal{E})$ is given
in Ref. \citep{chuang2009codeword}. A classical code $\mathcal{C}$
satisfying Eqs. (\ref{eq:cws classical condition 1}) to (\ref{eq:cws classical condition 3})
is a clique in the graph $G_{\mathcal{E}}=(N_{\mathcal{E}},E_{\mathcal{E}})$
with 
\begin{equation}
N_{\mathcal{E}}=\mathrm{GF}(2)^{n}\backslash(Cl_{G}(\mathcal{E})\cup D_{G}(\mathcal{E}))\label{eq:clique graph order}
\end{equation}
 and $E_{\mathcal{E}}$ defined by the classical error set $Cl_{G}(\mathcal{E})$
as outlined in Sec. \ref{subsec:Classical-codes}. That is, two nodes
$\boldsymbol{x}_{i},\boldsymbol{x}_{j}\in N_{\mathcal{E}}$ are connected
by an edge $\{\boldsymbol{x}_{i},\boldsymbol{x}_{j}\}\in E_{\mathcal{E}}$
if $\boldsymbol{x}_{i}+\boldsymbol{x}_{j}\notin Cl_{G}(\mathcal{E})$.
If $D_{G}(\mathcal{E})=\emptyset$, then for all $E\not\propto I\in\mathcal{E}$
it must be the case that $Cl_{G}(E)\neq\boldsymbol{0}$, and hence
$C_{E}=0$ in Eq. (\ref{eq:cws detection criteria}). Therefore, for
$E=E_{k}^{\dagger}E_{l}\in\mathcal{E}$ where $E_{k}E_{l}\in\mathcal{P}_{n}$
and $E_{k}\not\propto E_{l}$, it follows that $\langle W_{i}|E_{k}^{\dagger}E_{l}|W_{j}\rangle=0$.
That is, $\mathcal{Q}$ is pure if $D_{G}(\mathcal{E})=\emptyset$
\citep{chuang2009codeword,li2010structured}.

\subsection{Code bounds\label{subsec:Code-bounds}}

A simple, but relatively loose, upper bound on the dimension $K$
of an $n$-qubit code of distance $d$ is given by the quantum singleton
bound \citep{knill1997theory}
\begin{equation}
K\leq2^{n-2(d-1)}.
\end{equation}
A tighter limit on code size is given by the linear programming bound
\citep{calderbank1996quantum}. An $((n,K,d))$ code can exist only
if there are homogeneous polynomials $A(x,y)$, $B(x,y)$, and $S(x,y)$
such that
\begin{align}
A(1,0) & =1,\label{eq:LP1}\\
B(x,y) & =KA(\frac{x+3y}{2},\frac{x-y}{2}),\\
S(x,y) & =KA(\frac{x+3y}{2},\frac{y-x}{2}),\\
B(1,y)-A(1,y) & =O(y^{d}),\\
A(x,y) & \geq0,\\
B(x,y)-A(x,y) & \geq0,\\
S(x,y) & \geq0.\label{eq:LP7}
\end{align}
Here $C(x,y)\geq0$ means that the coefficients of the polynomial
$C$ are nonnegative, and $O(y^{d})$ is a polynomial in $y$ with
no terms of degree less than $d$. A pure $((n,K,d))$ code can exist
only if Eqs. (\ref{eq:LP1}) to (\ref{eq:LP7}) can be satisfied along
with
\begin{equation}
A(1,y)=1+O(y^{d}).\label{eq:LPpure}
\end{equation}
The linear programming bound is monotonic \citep{rains1998monotonicity},
meaning that if the constraints can be satisfied for some $K$, then
they can be satisfied for all lower code dimensions too. This monotonicity
holds even if $K$ is allowed to be a real number (rather than just
an integer). Following Ref. \citep{nebe2006self}, we define the real
number $K(n,d)$ as the largest $K>1$ for which Eqs. (\ref{eq:LP1})
to (\ref{eq:LP7}) can be satisfied. The purity conjecture of Ref.
\citep{calderbank1996quantum} states that if the linear programming
constraints hold for $K=K(n,d)$, then $A(1,y)=1+O(y^{d})$. The content
of this conjecture is simply that the linear programming bound for
pure codes is the same as for potentially impure codes. This conjecture
has been verified to hold for $n\leq100$ \citep{nebe2006self}.

For stabilizer codes, bounds on maximum $k$ are given in Table \ref{tab:stab bounds}
for $1\leq n\leq15$ and $2\leq d\leq5$. All lower bounds are given
by the best known stabilizer codes (these codes can be found at Ref.
\citep{Grassl:codetables}). The unmarked upper bounds are given by
the linear programming bound for $K=2^{k}$ (determined using YALMIP
\citep{Lofberg2004}). If the lower and upper bounds coincide, then
a single value is given; otherwise, they are separated by a dash.
In the cases marked ``A'', the $[[7,0,4]]$, $[[15,7,4]]$, and
$[[15,4,5]]$ codes that do not violate the linear programming bound
are excluded by arguments given in Sec. 7 of Ref. \citep{calderbank1996quantum}.
In the case marked ``B'', the $[[13,5,4]]$ code that does not violate
the linear programming bound is excluded by the argument of Ref. \citep{bierbrauer2009non}.
The entries marked ``C'' indicate cases where a code meeting the
bound must be impure (also outlined in Sec. 7 of Ref. \citep{calderbank1996quantum}).
An extended version of Table \ref{tab:stab bounds} for $n\leq256$
can be found at Ref. \citep{Grassl:codetables}.

\begin{table}
\caption{\label{tab:stab bounds}Bounds on the maximum $k$ of an $[[n,k,d]]$
stabilizer code for $1\protect\leq n\protect\leq15$ and $2\protect\leq d\protect\leq5$.}
\begin{tabular}{c|cccc}
$n\backslash d$ & $2$ & $3$ & $4$ & $5$\tabularnewline
\hline 
$1$ & $-$ & $-$ & $-$ & $-$\tabularnewline
$2$ & $0$ & $-$ & $-$ & $-$\tabularnewline
$3$ & $0$ & $-$ & $-$ & $-$\tabularnewline
$4$ & $2$ & $-$ & $-$ & $-$\tabularnewline
$5$ & $2$ & $1$ & $-$ & $-$\tabularnewline
$6$ & $4$ & $1^{C}$ & $0$ & $-$\tabularnewline
$7$ & $4$ & $1$ & $-^{A}$ & $-$\tabularnewline
$8$ & $6$ & $3$ & $0$ & $-$\tabularnewline
$9$ & $6$ & $3$ & $0$ & $-$\tabularnewline
$10$ & $8$ & $4$ & $2$ & $-$\tabularnewline
$11$ & $8$ & $5$ & $2$ & $1$\tabularnewline
$12$ & $10$ & $6$ & $4$ & $1^{C}$\tabularnewline
$13$ & $10$ & $7$ & $4^{B}$ & $1$\tabularnewline
$14$ & $12$ & $8$ & $6$ & $2-3$\tabularnewline
$15$ & $12$ & $9$ & $6^{A}$ & $3^{A}$\tabularnewline
\end{tabular}

\end{table}

Table \ref{tab:nonadditive bounds} gives the bounds on maximum $K$
for a potentially nonadditive $((n,K,d))$ code where $1\leq n\leq15$
and $2\leq d\leq5$. All upper bounds are from the linear programming
bound. The lower bounds marked ``A'' are from the family of nonadditive
$((2\alpha+1,3\times2^{2\alpha-3},2))$ codes of Ref. \citep{rains1999quantum}.
Those marked ``B'' are from the family of $((4\alpha+2\beta+3,M_{\alpha\beta},2))$
codes of Ref. \citep{smolin2007simple} where $\beta\in\{0,1\}$ and
\begin{equation}
M_{\alpha\beta}=\sum_{i=0}^{\alpha}\binom{4\alpha+2\beta+3}{2i+\beta}.
\end{equation}
The lower bounds marked ``C'' and ``D'' correspond to the $((9,12,3))$
and $((10,24,3))$ codes of Ref. \citep{yu2008nonadditive} and Ref.
\citep{yu2007graphical} respectively. All other lower bounds are
given by the best known stabilizer codes.

\begin{table}
\caption{\label{tab:nonadditive bounds}Bounds on the maximum size $K$ of
an $((n,K,d))$ code for $1\protect\leq n\protect\leq15$ and $2\protect\leq d\protect\leq5$.}
\begin{tabular}{c|cccc}
$n\backslash d$ & $2$ & $3$ & $4$ & $5$\tabularnewline
\hline 
$1$ & $-$ & $-$ & $-$ & $-$\tabularnewline
$2$ & $1$ & $-$ & $-$ & $-$\tabularnewline
$3$ & $1$ & $-$ & $-$ & $-$\tabularnewline
$4$ & $4$ & $-$ & $-$ & $-$\tabularnewline
$5$ & $^{A}6$ & $2$ & $-$ & $-$\tabularnewline
$6$ & $16$ & $2$ & $1$ & $-$\tabularnewline
$7$ & $^{A}24-26$ & $2-3$ & $0-1$ & $-$\tabularnewline
$8$ & $64$ & $8-9$ & $1$ & $-$\tabularnewline
$9$ & $^{A}96-112$ & $^{C}12-13$ & $1$ & $-$\tabularnewline
$10$ & $256$ & $^{D}24$ & $4-5$ & $-$\tabularnewline
$11$ & $^{B}386-460$ & $32-53$ & $4-7$ & $2$\tabularnewline
$12$ & $1,024$ & $64-89$ & $16-20$ & $2$\tabularnewline
$13$ & $^{B}1,586-1,877$ & $128-204$ & $16-40$ & $2-3$\tabularnewline
$14$ & $4,096$ & $256-324$ & $64-102$ & $4-10$\tabularnewline
$15$ & $^{B}6,476-7,606$ & $512-580$ & $64-150$ & $8-18$\tabularnewline
\end{tabular}
\end{table}

\section{Symmetric codes\label{sec:Symmetric-codes}}

An $((n,K,d))$ code must detect the set $\mathcal{E}^{d-1}$ as defined
in Eq. (\ref{eq:pauli basis}). Note that $\mathcal{E}^{1}$, and
hence $\mathcal{E}^{d-1}$ more generally, is invariant under any
permutation of the Pauli matrices $X$, $Y$, and $Z$ on any subset
of qubits. As a result of this symmetry, we call $((n,K,d))$ codes
symmetric codes. Furthermore, as outlined in Sec. \ref{subsec:Quantum-channels-and},
this symmetry means that if some code $\mathcal{Q}$ detects $\mathcal{E}^{d-1}$,
then so does any equivalent code $\mathcal{Q}'$. It is therefore
sufficient to consider only standard form codes when attempting to
construct an optimal symmetric CWS code. Furthermore, we need only
consider standard form codes based on representatives from different
elements of $\mathcal{L}_{n}$. However, as outlined in Sec. \ref{subsec:Undirected-graphs}
the size of $\mathcal{L}_{n}$ appears to grow exponentially, and
it has only been enumerated for $n\leq12$. Furthermore, constructing
an optimal classical code for a given graph by finding a maximum clique
is NP-hard as mentioned in Sec. \ref{subsec:Undirected-graphs}. In
this section we explore methods of code construction that address
these two obstacles.

\subsection{Distance two codes}

\begin{figure}
\includegraphics[scale=0.55]{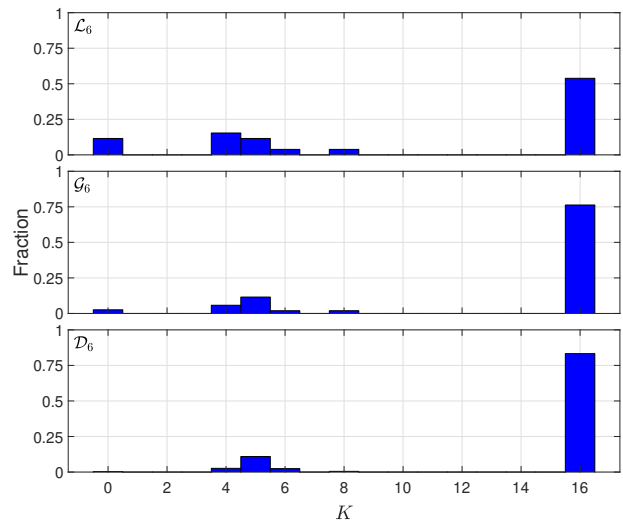}\caption{\label{n6d2 hist}Code size distributions for non-LC-isomorphic, nonisomorphic,
and distinct graphs in the case of $n=6$ and $d=2$.}
\end{figure}

First we consider distance two codes of even length. As outlined in
Tables \ref{tab:stab bounds} and \ref{tab:nonadditive bounds}, there
are even length stabilizer codes with $k=n-2$ that saturate the singleton
bound for $n\leq14$. In fact, there are stabilizer codes that saturate
the bound for all even $n$ \citep{rains1999quantum}. Despite this,
there is still some insight to be gained from constructing CWS codes
with these parameters. For $n\leq10$, it is feasible to exhaustively
search $\mathcal{L}_{n}$ (that is, to construct a code based on a
representative of each element of $\mathcal{L}_{n}$). Using the code
size distribution over $\mathcal{L}_{n}$, it is possible to determine
the distributions over $\mathcal{G}_{n}$ and $\mathcal{D}_{n}$ by
counting the number of nonisomorphic and distinct graphs respectively
in each element of $\mathcal{L}_{n}$ (see Sec. \ref{subsec:Undirected-graphs}).
As an example, the code size distributions for $n=6$ are shown in
Fig. \ref{n6d2 hist}. It can be seen that over $50\%$, $75\%$,
and $80\%$ of elements of $\mathcal{L}_{6}$, $\mathcal{G}_{6}$,
and $\mathcal{D}_{6}$ respectively yield optimal $K=16$ codes. The
fraction of elements of $\mathcal{L}_{n}$, $\mathcal{G}_{n}$, and
$\mathcal{D}_{n}$ that yield optimal codes for even $2\leq n\leq10$
is shown in Table \ref{d=00003D2 even n optimal code dist}. For $2\leq n\leq6$,
the clique graphs generated are small enough for maximum cliques to
be found using the exact algorithm of Ref. \citep{konc2007improved}.
For $n\geq8$, we have resorted to using the approximate PLS algorithm
due to the larger clique graphs. We have allowed the PLS algorithm
$100$ attempts, each of which used a maximum of $1,000$ selections
(these are the default PLS parameters that we have employed). As a
result of having used an approximate clique finding algorithm, the
values given in the $n=8$ and $n=10$ rows of Table \ref{d=00003D2 even n optimal code dist}
are a lower bounds. It can be seen that in each case, the fraction
of elements in $\mathcal{D}_{n}$ yielding an optimal code is greater
than that of $\mathcal{G}_{n}$, which in turn is greater than that
of $\mathcal{L}_{n}$. Furthermore, increasing $n$ increases the
fraction of optimal codes in all cases. In particular, by $n=10$
over $98\%$ of distinct graphs yield a code with an optimal $K=256$.
This trend suggests that for larger $n$, we are highly likely to
find an optimal code even if we use a randomly selected graph. This
goes some way to explaining the results of Ref. \citep{yen2009optimal},
where cycle graphs were shown to give optimal codes for even $n\leq12$. 

\begin{table}
\caption{\label{d=00003D2 even n optimal code dist}The fraction of elements
of $\mathcal{L}_{n}$, $\mathcal{G}_{n}$, and $\mathcal{D}_{n}$
that yield optimal $K=2^{n-2}$ codes for even $n\protect\leq10$
and $d=2$. The values given for $n=8$ and $n=10$ are lower bounds.}
\begin{tabular}{c|ccc}
$n$ & $\mathcal{L}_{n}$ & $\mathcal{G}_{n}$ & $\mathcal{D}_{n}$\tabularnewline
\hline 
$2$ & $0.500$ & $0.500$ & $0.500$\tabularnewline
$4$ & $0.500$ & $0.636$ & $0.641$\tabularnewline
$6$ & $0.539$ & $0.763$ & $0.833$\tabularnewline
$8$ & $0.643$ & $0.909$ & $0.938$\tabularnewline
$10$ & $0.815$ & $0.977$ & $0.981$\tabularnewline
\end{tabular}
\end{table}

\begin{figure}
\includegraphics[scale=0.55]{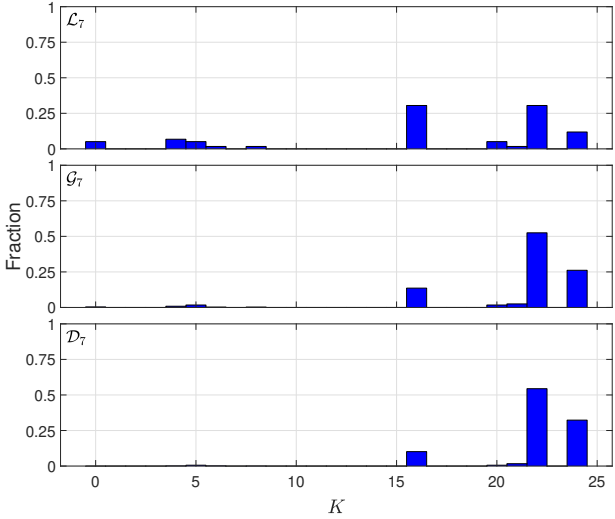}\caption{\label{n7d2 hist}Code size distributions for non-LC-isomorphic, nonisomorphic,
and distinct graphs in the case of $n=7$ and $d=2$.}
\end{figure}

The case of odd $n$ is somewhat more interesting. Here, as shown
in Ref. \citep{rains1999quantum}, the linear programming bound reduces
to 

\begin{equation}
K\leq2^{n-2}(1-\frac{1}{n-1}).\label{eq:odd n d2 bound}
\end{equation}
Stabilizer codes cannot saturate this bound and are restricted to
$k\leq n-3$. Again, we can construct codes based on an exhaustive
search of $\mathcal{L}_{n}$ for $n\leq11$. For $n=3$, a single
element of $\mathcal{L}_{3}$ yields an optimal $K=1$ code. Similarly,
a single element of $\mathcal{L}_{5}$ yields a code with $K=6$,
which matches the size of the optimal code given in Refs. \citep{rains1997nonadditive,rains1999quantum}.
For $n=7$, there is more of a spread in the code sizes as shown in
Fig. \ref{n7d2 hist}. It can be seen that a large number of graphs
yield codes with $K=16$ or $K=22$, which match the size of an optimal
stabilizer code and the code of Ref. \citep{smolin2007simple} respectively.
Furthermore, there are seven elements of $\mathcal{L}_{7}$ that yield
codes with $K=24$, which match the size of the code of Ref. \citep{rains1999quantum}.
No graphs yield codes with $K=25$ or $K=26$, despite such codes
not being excluded by the linear programming bound.

\begin{figure}
\includegraphics[scale=0.55]{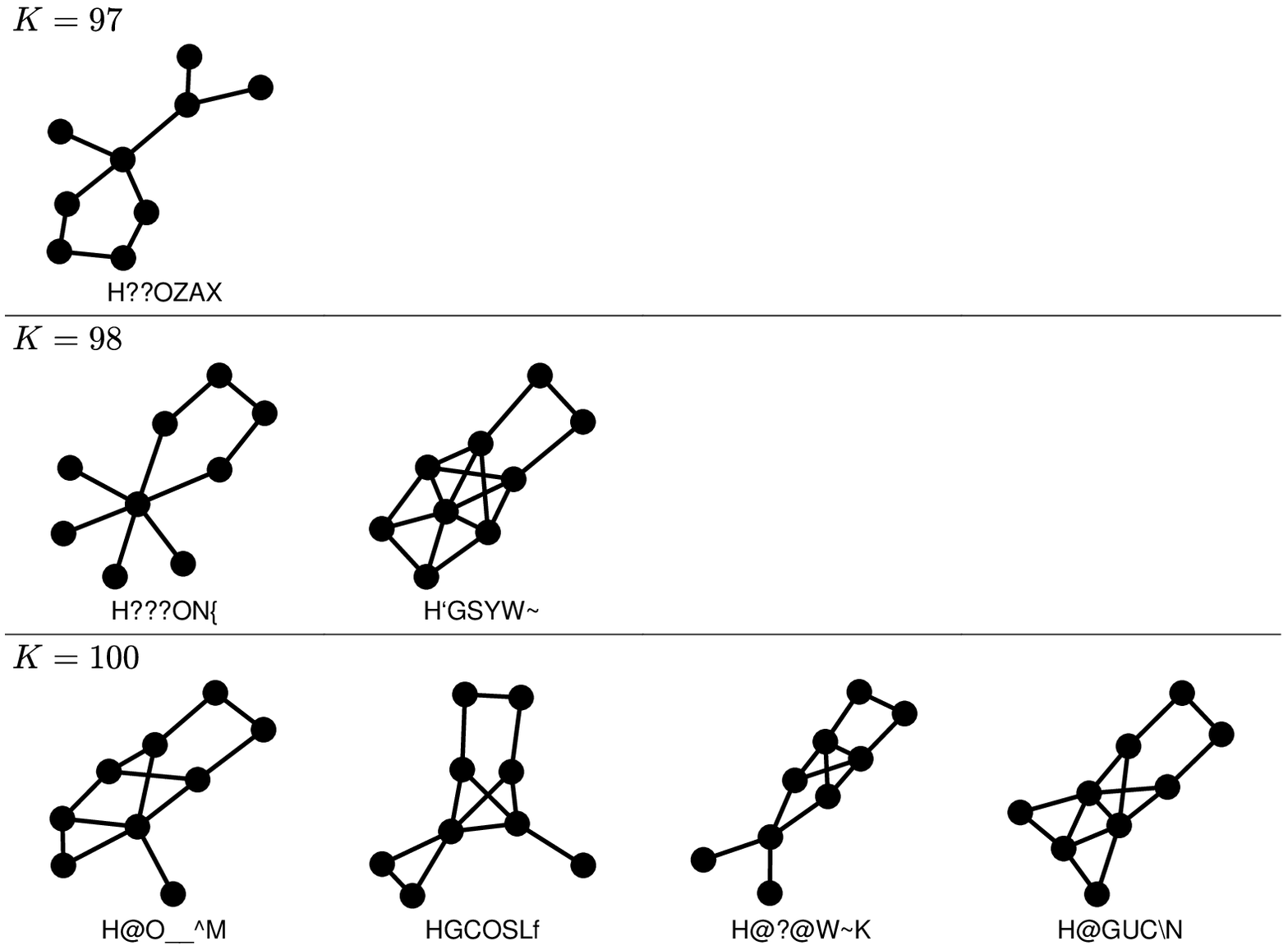}\caption{\label{n9d2 graphs}Non-LC-isomorphic graphs that yield $((9,97\protect\leq K\protect\leq100,2))$
codes.}

\end{figure}

For $n=9$, an exhaustive search of $\mathcal{L}_{9}$ is still feasible;
however, we have done so using the PLS clique finder, and as such
there may exist larger CWS codes than the ones reported here. Similar
to the $n=7$ case, the majority of graphs gave codes with $K=64$,
$K=93$, or $K=96$, which match the size of an optimal stabilizer
code, the code of Ref. \citep{smolin2007simple}, and the code of
Ref. \citep{rains1999quantum} respectively. However, we have also
found seven elements of $\mathcal{L}_{9}$ that yield codes with $K\geq97$.
To increase the likelihood that we have found maximum size codes for
these seven graphs, we have repeated the clique search for each of
them using $10,000$ attempts. This has resulted in one $K=97$ code,
two $K=98$ codes, and four $K=100$ codes. Representatives of the
elements of $\mathcal{L}_{n}$ that yielded these codes are shown
in Fig. \ref{n9d2 graphs}. Note that we do not label the nodes as
isomorphic graphs yield equivalent codes. Given below each of the
drawings is the graph in graph6 format (see Ref. \citep{mckay2013nauty}
for details). While these $K\geq97$ codes are larger than any previously
known codes, they do not saturate the linear programming bound of
$K=112$. A classical code for each of these graphs is given in the
supplementary material (this is the case for all codes presented in
this paper).

For $n=11$, we have performed an exhaustive search of $\mathcal{L}_{11}$
with an increased $10,000$ PLS selections to account for the larger
cliques. Here we have mostly obtained codes with $K=256$, $K=384$,
or $K=386$, which match the size of an optimal stabilizer code, the
code of Ref. \citep{rains1999quantum}, and the code of Ref. \citep{smolin2007simple}
respectively. We have also found $413$ elements of $\mathcal{L}_{11}$
that yield codes with $K\geq387$. As for the $n=9$ case, we have
repeated the clique search for these graphs using $10,000$ attempts.
The resulting code size distribution is given in Table \ref{n11d2 table}.
Representatives of elements of $\mathcal{L}_{11}$ that yield codes
with $K\geq406$ are shown in Fig. \ref{n11d2 graphs} (the remaining
graphs are included in the supplementary material). Again, while these
are the largest codes known, they do not saturate the linear programming
bound of $K=460$. As $\mathcal{L}_{n}$ has not been enumerated for
$n\geq13$, we cannot continue this exhaustive search procedure for
higher $n$. Any (nonexhaustive) search of $\mathcal{G}_{n}$ or $\mathcal{D}_{n}$
is also impractical for $n\geq13$ due to the large clique graphs
produced, which both makes the clique search slow and reduces the
likelihood that the clique found is of maximum size.

\begin{table}
\caption{\label{n11d2 table}Number of elements $N_{K}$ of $\mathcal{L}_{11}$
that gave codes of given size $K$ with $d=2$.}
\begin{tabular}{c|ccccccccccccc}
$K$ & $387$ & $388$ & $389$ & $390$ & $391$ & $392$ & $398$ & $400$ & $402$ & $404$ & $406$ & $408$ & $416$\tabularnewline
\hline 
$N_{K}$ & $51$ & $11$ & $1$ & $1$ & $2$ & $54$ & $2$ & $207$ & $1$ & $74$ & $1$ & $6$ & $2$\tabularnewline
\end{tabular}
\end{table}

\begin{figure}
\includegraphics[scale=0.55]{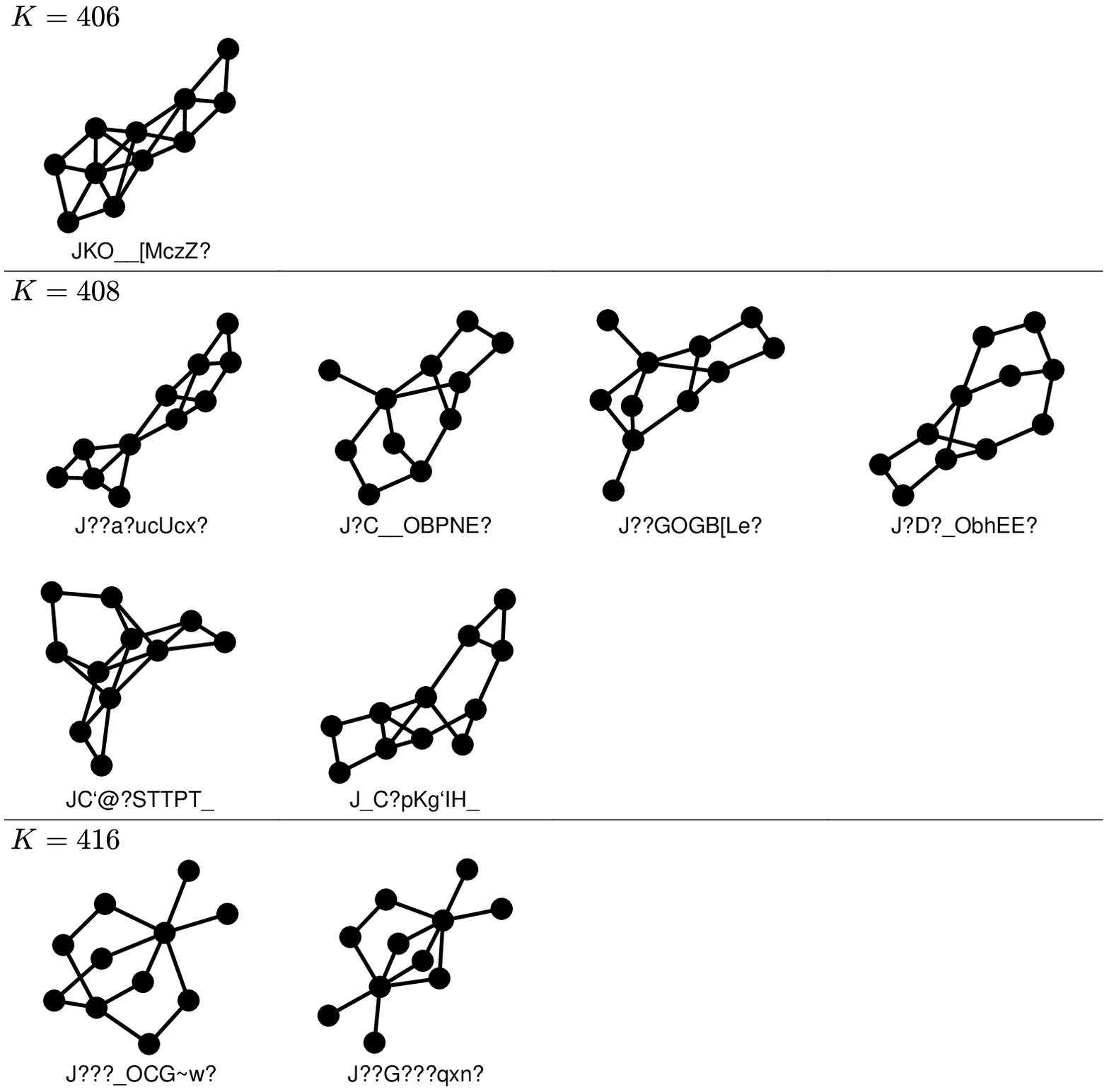}\caption{\label{n11d2 graphs}Non-LC-isomorphic graphs that yield $((11,406\protect\leq K\protect\leq416,2))$
codes.}
\end{figure}

\begin{figure}
\includegraphics[scale=0.55]{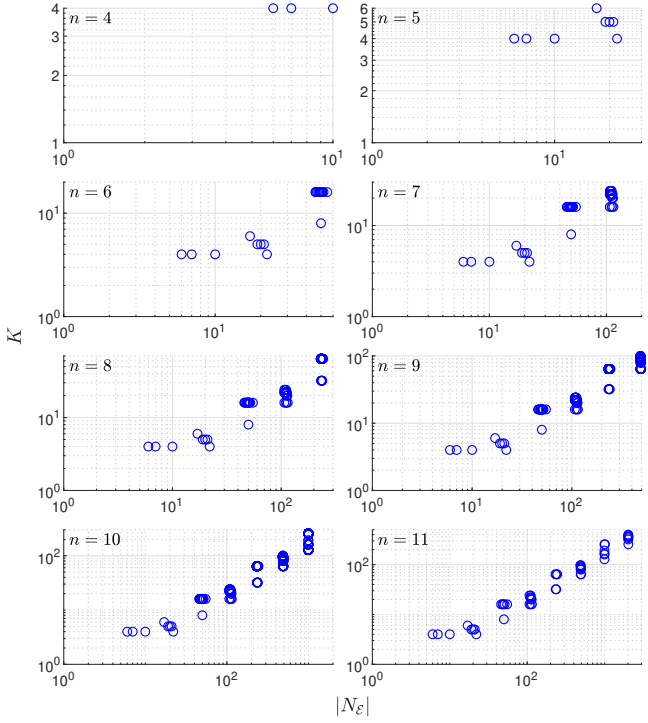}\caption{\label{fig:d2scatters}Clique graph order vs code size for codes with
$4\protect\leq n\protect\leq11$ and $d=2$.}

\end{figure}

Figure \ref{fig:d2scatters} shows the relationship between code size
and clique graph order $|N_{\mathcal{E}}|$ for $4\leq n\leq11$.
It can be seen that the data is clustered by clique graph order; furthermore,
in each case, the graphs yielding the largest codes belong to the
highest $|N_{\mathcal{E}}|$ cluster. This clustering behavior can
be explained by considering Eq. (\ref{eq:clique graph order}), which
gives
\begin{equation}
|N_{\mathcal{E}}|=2^{n}-|Cl_{G}(\mathcal{E})|-|D_{G}(\mathcal{E})|.\label{eq:clique graph size}
\end{equation}
It follows from Eq. (\ref{eq:degenerate set}) that $\mathrm{GF}(2)^{n}\backslash D_{G}(\mathcal{E})$
is the annihilator of $\mathcal{E}'=\{E\in\mathcal{E}:Cl_{G}(E)=0\}$
and is therefore a subspace of $\mathrm{GF}(2)^{n}$. If $\dim(\mathrm{GF}(2)^{n}\backslash D_{G}(\mathcal{E}))=r\leq n$,
then $|D_{G}(\mathcal{E})|=2^{n}-2^{r}$, which gives $|N_{\mathcal{E}}|=2^{r}-|Cl_{G}(\mathcal{E})|$.
The clusters therefore correspond to different values of $r$. The
codes in the highest $|N_{\mathcal{E}}|$ cluster are pure as they
have $D_{G}(\mathcal{E})=\emptyset$. That this cluster contains codes
of maximum size is not entirely surprising in light of the purity
conjecture outlined in Sec. \ref{subsec:Code-bounds}.

\subsection{Distance three codes}

Distance three codes are of practical interest as they allow for the
correction of an arbitrary single-qubit error. For $n\leq11$, we
can exhaustively search $\mathcal{L}_{n}$ in the same way as we have
for the distance two codes of the previous section. There are one
and two elements of $\mathcal{L}_{5}$ and $\mathcal{L}_{6}$ respectively
that give optimal $K=2$ codes (note that all $K=2$ CWS codes are
additive \citep{chuang2009codeword}). Similarly, there are $18$
elements of $\mathcal{L}_{7}$ that yield $K=2$ codes. As has been
previously shown in Ref. \citep{chuang2009codeword}, although the
linear programming bound does not exclude them, there are no $((7,3,3))$
CWS codes. There are six elements of $\mathcal{L}_{8}$ that give
$K=8$ codes. No elements yield a $K=9$ code, despite such a code
not being excluded by the linear programming bound. There are eight
elements of $\mathcal{L}_{9}$ that yield $K=12$ codes, which match
the size of the code presented in Ref. \citep{yu2008nonadditive}.
Again, no elements yield a $K=13$ code, despite such a code not being
excluded by the linear programming bound. An exhaustive search of
$\mathcal{L}_{10}$ has previously been performed in Ref. \citep{yu2007graphical},
where it was shown that a single element yields an optimal $K=24$
code. We have exhaustively searched $\mathcal{L}_{11}$ using the
PLS clique finder. This has yielded $13,709$ $K=32$ codes, which
match the size of an optimal stabilizer code. No larger codes were
found, which is somewhat surprising given that the linear programming
bound is $K=53$.

\begin{figure}
\includegraphics[scale=0.55]{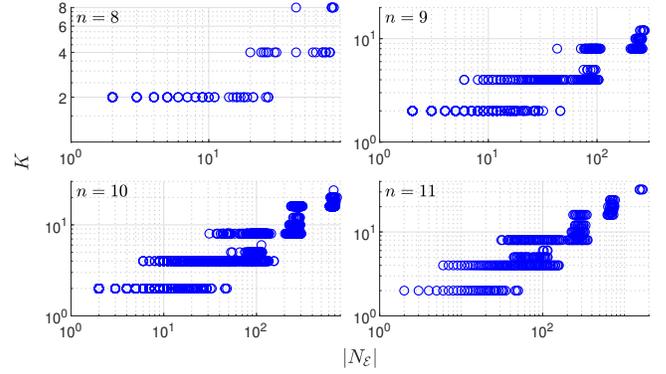}\caption{\label{fig:d3scatters}Clique graph order vs code size for codes with
$8\protect\leq n\protect\leq11$ and $d=3$.}
\end{figure}

Figure \ref{fig:d3scatters} shows the relationship between code size
and clique graph order for distance three codes with $8\leq n\leq11$.
It can be seen that there is greater spread within the clusters compared
to the distance two case of Fig. \ref{fig:d2scatters}. According
to Eq. (\ref{eq:clique graph size}), this can be attributed to an
increased variance in the size of $Cl_{G}(\mathcal{E})$. Despite
this increased variation, the graphs yielding the best codes belong
to the highest $|N_{\mathcal{E}}|$ cluster in all four cases. Importantly,
the best codes are not necessarily given by the graphs with the highest
clique graph order within this cluster. For example, in the $n=10$
case, the highest clique graph cluster contains graphs with $613\leq|N_{\mathcal{E}}|\le739$,
while the graph yielding the $K=24$ code only has $|N_{\mathcal{E}}|=679$.

\begin{figure}
\includegraphics[scale=0.55]{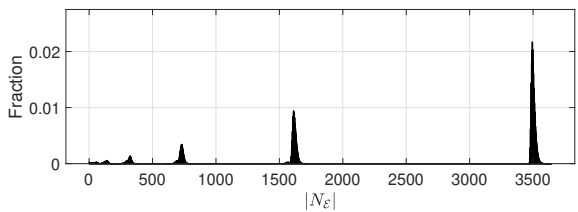}\caption{\label{n12d3Vhist}Clique graph order distribution over $\mathcal{L}_{12}$
for codes with with $d=3$.}

\end{figure}

For $n=12$, the size of $\mathcal{L}_{12}$ makes an exhaustive search
somewhat prohibitive. We can reduce the search space somewhat by considering
the distribution of clique graph sizes as shown in Fig. \ref{n12d3Vhist}.
Note that by using Eq. (\ref{eq:clique graph size}), $|N_{\mathcal{E}}|$
can be computed without actually constructing the clique graph. Our
previous observations regarding the relationship between code size
and clique graph order suggest that graphs yielding the best codes
are highly likely to be found in the $|N_{\mathcal{E}}|>3,000$ cluster.
We have randomly selected $50,000$ of the $663,039$ elements of
$\mathcal{L}_{12}$ in this cluster and constructed a code for each
using the PLS clique finder. This has yielded $6,325$ codes with
$K=64$, which match the size of an optimal stabilizer code. No larger
codes were found, despite the linear programming bound not excluding
codes with up to $K=89$. We have not pursued searches for $n\geq13$
codes as while the clique graphs produced are smaller than in the
$d=2$ case, they are still large enough for maximum clique searches
to be unreliable.

\subsection{Distance four codes\label{subsec:Distance-four-codes}}

\begin{figure}
\includegraphics[scale=0.55]{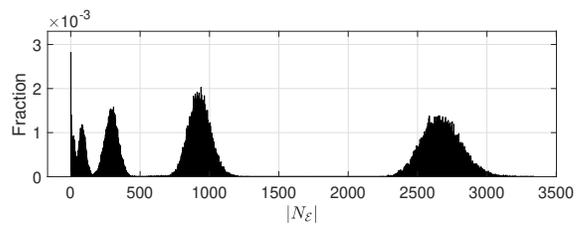}\caption{\label{n13d4Vhist}Clique graph order distribution over $\mathcal{D}_{13}$
for codes with with $d=4$.}
\end{figure}

For $d=4$, we are able to perform exhaustive searches of $\mathcal{L}_{n}$
for $n\leq12$. There are one, five, and eight elements of $\mathcal{L}_{6}$,
$\mathcal{L}_{8}$, and $\mathcal{L}_{9}$ respectively that yield
optimal $K=1$ codes. As expected, no elements of $\mathcal{L}_{7}$
give a nontrivial code (note that a $K=1$ CWS code is a stabilizer
state and hence pure; such $[[n,0,d]]$ codes have previously been
classified in \citep{danielsen2006classification}). There are $10$
and $3,060$ elements of $\mathcal{L}_{10}$ and $\mathcal{L}_{11}$
respectively that give $K=4$ codes, which match the size of an optimal
stabilizer code. No elements yield larger codes, despite the linear
programming bound not excluding codes with up to $K=5$ and $K=7$
respectively. Unlike the $d=3$ case, an exhaustive search of $\mathcal{L}_{12}$
is feasible for $d=4$ due to the smaller clique graphs. However,
the clique graphs are still large enough that we have resorted to
using the PLS clique finding algorithm. This search has yielded $1,482$
codes with $K=16$, which match the size of an optimal stabilizer
code. No larger codes were found, despite the linear programming bound
not excluding codes with up to $K=20$. The smaller clique graph sizes
in the $d=4$ case also make searching for codes with $n=13$ and
$n=14$ feasible. For $n=13$, we have randomly selected $100,000$
graphs from $\mathcal{D}_{13}$ to estimate the clique graph size
distribution as shown in Fig. \ref{n13d4Vhist}. $41,458$ of these
graphs belong to the $|N_{\mathcal{E}}|>2,000$ cluster. Of these,
one yielded a $K=18$ code, which is larger than an optimal $K=16$
stabilizer code.

\begin{figure}
\includegraphics[scale=0.55]{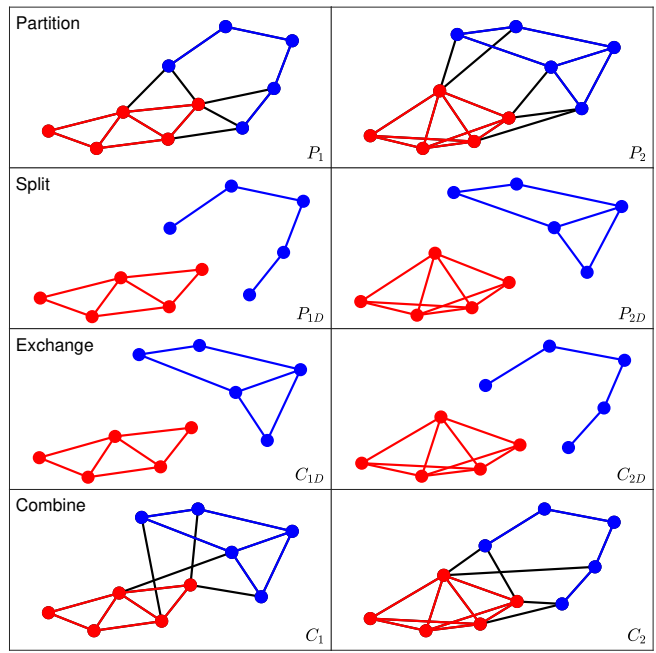}\caption{\label{spectral crossover fig}Spectral crossover example for $n=10$
graphs. Each parent graph is split into two fragments according to
a spectral bisection. These fragments are then exchanged and combined
to form two child graphs.}
\end{figure}

To find more $n=13$ codes with $K>16$, we want a more reliable way
of generating graphs that yield a large clique graph. That is, we
wish to search $\mathcal{D}_{n}$ for graphs yielding a large clique
graph in a way that is more efficient than a random search. We have
found a genetic algorithm to be effective in this respect. There are
a number of ways we could implement mutation and crossover in this
algorithm. For mutation, we first select two nodes in the child graph
at random. If these two nodes are not connected by an edge, then one
is added; otherwise, if they are connected by an edge, then it is
removed. If we represent the parent graphs as bit strings, then we
can use standard single-point, two-point, or uniform crossover. One
way to achieve this is to convert the upper triangular component of
a parent adjacency matrix to a bit string row by row. Alternatively,
we can use a graph based approach. However, the method of Ref. \citep{globus1999automatic}
outlined in Sec. \ref{subsec:Genetic-algorithms} is not appropriate
for searching $\mathcal{D}_{n}$ as it is not guaranteed to produce
child graphs with $n$ nodes. Furthermore, as previously mentioned,
it tends to remove an unnecessarily large number of edges when splitting
the parent graphs into two fragments. To address these issues, we
propose splitting the parent graphs using a spectral bisection. In
particular, the nodes of a parent graph $P$ are bisected into the
sets $N_{1}$ and $N_{2}$, which define the fragments $F_{1}=P[N_{1}]$
and $F_{2}=P[N_{2}]$. A fragment is then exchanged between each parent
to form two disconnected children that are then connected following
the method of Ref. \citep{globus1999automatic}. An example of this
procedure on two $n=10$ graphs is shown in Fig. \ref{spectral crossover fig}. 

\begin{figure}
\includegraphics[scale=0.55]{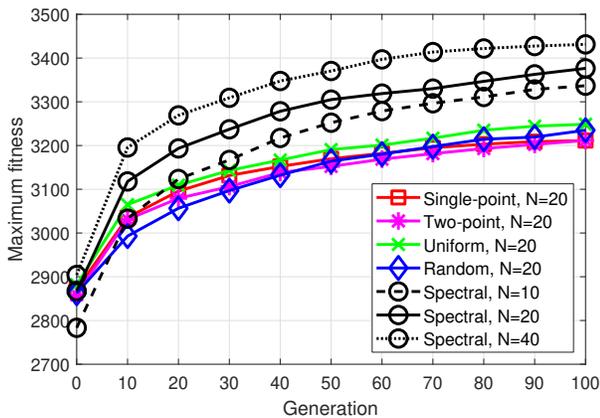}\caption{\label{crossover comparison}Comparison of crossover methods for $n=13$,
$d=4$ codes. The vertical axis shows the fitness (the clique graph
order $|N_{\mathcal{E}}|$) of the highest fitness element of the
child population averaged over $100$ genetic algorithm instances.}
\end{figure}

We have run $100$ genetic algorithm instances using each of the potential
crossover methods to compare their performance. In each instance,
we have used a population size of $N=20$, $100$ generations, a crossover
probability of $p_{c}=0.9$, a mutation probability of $p_{m}=0.1$,
and a tournament size of $10$. We have also incorporated elitist
selection, with the fittest two parent graphs (that is, the two that
yield the largest clique graphs) being added to the child population
at the start of each generation. The average order of the highest
order clique graph yielded in each generation is shown in Fig. \ref{crossover comparison}.
It can be seen that single-point, two-point, and uniform crossover
(with $p_{e}=0.5$) all exhibit similar performance. However, their
performance is also matched by random crossover, where the two children
are simply selected at random from $\mathcal{D}_{n}$ with no input
from the parents. As such, the increase in fitness with successive
generations when using these crossover methods is simply due to the
selection pressure of the genetic algorithm. It can also be seen that
spectral crossover gives significantly better performance than all
other methods. We have also tested the effect of population size when
using spectral crossover. In particular, we have tested population
sizes of $N=10$ and $N=40$ in addition to the previously considered
$N=20$ case. We have used a tournament size of half the population
size in each case and left all other parameters unchanged. It can
be seen in Fig. \ref{crossover comparison} that, as expected, increasing
the population size increases the average maximum fitness. With clique
graph order only serving as an indicator of code size, it is not essential
for the genetic algorithm to find graphs that yield the absolute largest
clique graphs. In fact, as was seen in the $n=10$, $d=3$ case, focusing
solely on such graphs may mean that we miss the best code(s). With
this in mind, we have found using $50$ generations and a population
size of $N=10$ to be a good compromise. Using a modest population
size and number of generations is also favorable from a run time perspective
as determining $|N_{\mathcal{E}}|$ becomes more computationally expensive
with increasing code length and/or distance (both of which serve to
increase the size of the error set).

\begin{figure}
\includegraphics[scale=0.55]{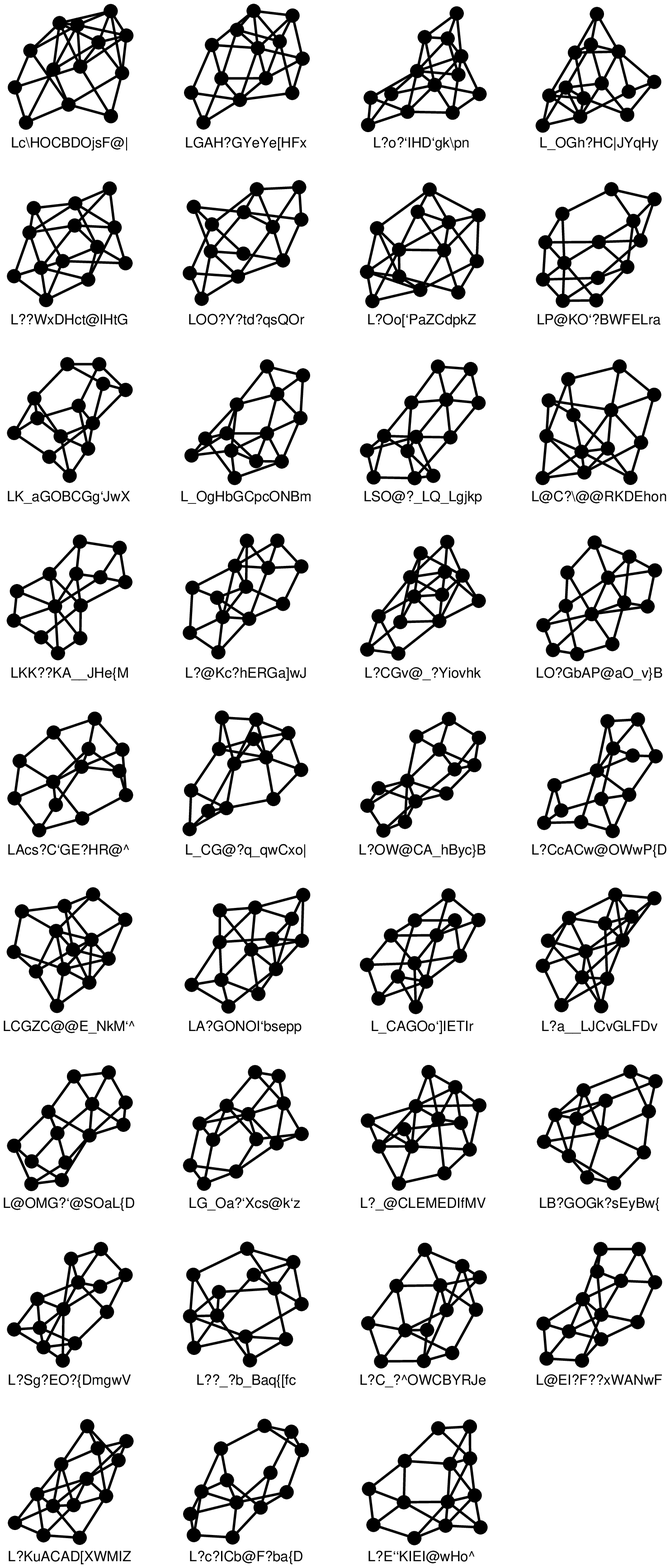}\caption{\label{n13d4 K18 graphs}Non-LC-isomorphic graphs that yield $((13,18,4))$
codes.}
\end{figure}

\begin{figure}
\includegraphics[scale=0.55]{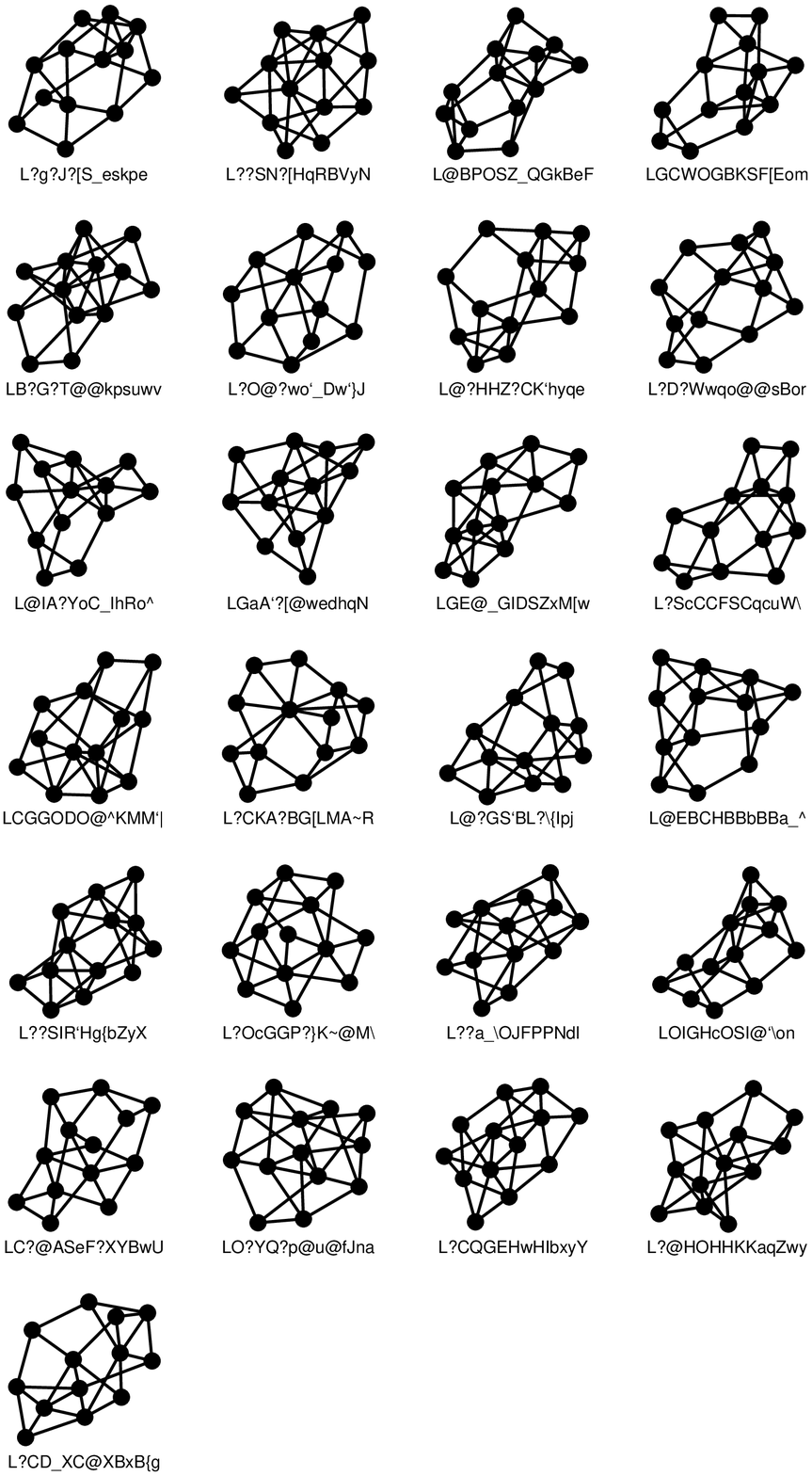}\caption{\label{n13d4 K20 graphs}Non-LC-isomorphic graphs that yield $((13,20,4))$
codes.}
\end{figure}

The genetic algorithm we have outlined is quite greedy/exploitative.
To make our search more explorative, we run a large number of genetic
algorithm instances, with a code being constructed from the fittest
graph found by each instance. For $n=13$, we have run $50,000$ such
instances, of which $352$ yielded a $K=18$ code and a further $175$
gave a $K=20$ code. The graphs that yielded codes with $K=18$ and
$K=20$ belong to $35$ and $25$ different elements of $\mathcal{L}_{13}$
respectively. A representative from each of these elements is shown
in Figs. \ref{n13d4 K18 graphs} and \ref{n13d4 K20 graphs}. Note
that the graphs shown are not necessarily the exact ones found using
the genetic algorithm; they are LC-equivalent graphs that can be drawn
clearly using the force-directed layout method of Ref. \citep{fruchterman1991graph}.
While these $K=18$ and $K=20$ codes are larger than any previously
known codes, they do not saturate the linear programming bound of
$K=40$. We have also run $50,000$ instances of the genetic algorithm
for $n=14$. $65$ of these instances have yielded $K=64$ codes,
which match the size of an optimal stabilizer code. We have not found
any codes with $K>64$, despite the linear programming bound not excluding
codes with up to $K=102$.

\subsection{Distance five codes}

For $d=5$, one and five elements of $\mathcal{L}_{11}$ and $\mathcal{L}_{12}$
respectively yield optimal $K=2$ codes. For $13\leq n\leq15$ we
have run $50,000$ genetic algorithm instances. $46,978$ instances
yielded a $K=2$ code for $n=13$, $452$ instances yielded a $K=4$
code for $n=14$, and $14$ instances yielded a $K=8$ code for $n=15$.
No larger codes were found, despite the linear programming bound being
$K=3$, $K=10$, and $K=18$ respectively. Note that the existence
of a $((13,3,5))$ CWS code has already been excluded in Ref. \citep{chuang2009codeword}
by the same argument that excluded the $((7,3,3))$ code.

\section{Asymmetric codes\label{sec:Asymmetric-codes}}

A channel of physical interest is the amplitude damping channel
\begin{equation}
\rho\rightarrow A_{0}\rho A_{0}^{\dagger}+A_{1}\rho A_{1}^{\dagger},
\end{equation}
where
\begin{equation}
A_{0}=\left(\begin{array}{cc}
1 & 0\\
0 & \sqrt{1-\gamma}
\end{array}\right),\,A_{1}=\left(\begin{array}{cc}
0 & \sqrt{\gamma}\\
0 & 0
\end{array}\right).
\end{equation}
It can be shown \citep{grassl2014quantum,gottesman1997stabilizer,fletcher2008channel}
that a sufficient condition for correcting a single amplitude damping
error is the ability to detect
\begin{equation}
\mathcal{E}^{\{1\}}=\{I,X_{i},Y_{i},Z_{i},X_{i}X_{j},X_{i}Y_{j},Y_{i}Y_{j}\},
\end{equation}
where $1\leq i,j\leq n$. This is not a necessary condition for correcting
an amplitude damping error. In fact, a code detecting $\mathcal{E}^{\{1\}}$
can also correct a single $A_{1}^{\dagger}$ error \citep{jackson2016codeword}.
A code can correct $t$ amplitude damping errors if it can detect
$\mathcal{E}^{\{t\}}$, which is comprised of all $t$-fold combinations
of elements from $\mathcal{E}^{\{1\}}$. $\mathcal{E}^{\{t\}}$ is
a subset of $\mathcal{E}^{2t}$, which is the set of errors that must
be detected to guarantee the ability to correct an arbitrary weight
$t$ error. As a result, there is potential for constructing codes
correcting $t$ amplitude damping errors that are larger than those
correcting $t$ arbitrary errors. For example, the stabilizer codes
presented in Ref. \citep{gottesman1997stabilizer} detect $\mathcal{E}^{\{1\}}$
and have the parameters given in Table \ref{single ad table} (these
values are taken from Ref. \citep{shor2011high}). In all but the
$n=8$ case, these codes are larger than the size of an optimal $d=3$
stabilizer code as given in Table \ref{tab:stab bounds}. An exhaustive
search for CWS codes detecting $\mathcal{E}^{\{1\}}$ has been performed
in Ref. \citep{jackson2016codeword} for $5\leq n\leq9$. The size
of these codes is also given in Table \ref{single ad table}, where
they can be seen to be larger than the stabilizer codes for $n=8$
and $n=9$. Other nonadditive codes have also been constructed that
can correct a single amplitude damping error \citep{lang2007nonadditive,shor2011high};
however, they are not directly comparable as they do so in a way that
does not guarantee the detection of $\mathcal{E}^{\{1\}}$ (that is,
they cannot correct an $A_{1}^{\dagger}$ error).

\begin{table}
\caption{\label{single ad table}Size of stabilizer codes presented in Ref.
\citep{gottesman1997stabilizer} and CWS codes presented in Ref. \citep{jackson2016codeword}
that detect $\mathcal{E}^{\{1\}}$.}
\begin{tabular}{c|cccccccccccc}
$n$ & $4$ & $5$ & $6$ & $7$ & $8$ & $9$ & $10$ & $11$ & $12$ & $13$ & $14$ & $15$\tabularnewline
\hline 
Stabilizer & $1$ & $2$ & $4$ & $8$ & $8$ & $16$ & $32$ & $64$ & $128$ & $256$ & $512$ & $1,024$\tabularnewline
CWS & $-$ & $2$ & $4$ & $8$ & $10$ & $20$ & $-$ & $-$ & $-$ & $-$ & $-$ & $-$\tabularnewline
\end{tabular}
\end{table}

$\mathcal{E}^{\{t\}}$ is not invariant under all possible Pauli matrix
permutations. As such, two LC-equivalent CWS codes need not correct
the same number of amplitude damping errors. This means that considering
standard form codes based on different elements of $\mathcal{L}_{n}$
no longer constitutes an exhaustive search of all CWS codes. However,
as suggested in Ref. \citep{jackson2016codeword}, a search of $\mathcal{L}_{n}$
can be made exhaustive by performing it for every LC-equivalent error
set of the form $U^{\dagger}\mathcal{E}^{\{t\}}U$. These sets are
versions of $\mathcal{E}^{\{t\}}$ with $X$, $Y$, and $Z$ errors
permuted on some set of qubits. If $\mathcal{E}^{\{t\}}$ exhibited
no symmetries under such permutations, then there would be $6^{n}$
such sets. However, as $\mathcal{E}^{\{t\}}$ is invariant under the
permutation $X\leftrightarrow Y$ on any subset of qubits, this number
is reduced to $3^{n}$. Unfortunately, an exhaustive search is not
practical for codes with $n\geq10$ as even for $n=10$ there are
$3^{10}|\mathcal{L}_{10}|=235,605,510$ cases to test. In this section,
we build on our code construction methods to address this increase
in the size of the search space.

\subsection{Single amplitude damping error}

To construct new codes for the amplitude damping channel with $n\geq10$,
we first consider $n\leq9$ to determine what types of codes match
the bounds provided in Ref. \citep{jackson2016codeword}. Initially,
we restrict consideration to standard form codes that detect $\mathcal{E}^{\{1\}}$.
As $\mathcal{E}^{\{1\}}$ (and $\mathcal{E}^{\{t\}}$ more generally)
is invariant under a permutation of qubit labels, it is sufficient
to consider one representative from each element of $\mathcal{G}_{n}$.
The first column of Table \ref{tab:AD table} shows the number of
elements of $\mathcal{G}_{n}$ for $5\leq n\leq9$ that yield optimal
standard form CWS codes. Note that the value given for $n=9$ is a
lower bound as we have used the PLS clique finder in this case. It
can be seen that while we are able to construct optimal codes for
$5\leq n\leq7$, we are unable to do so for $n=8$ and $n=9$. To
remedy this, we consider the LC-equivalent error sets
\begin{equation}
\mathcal{E}_{XZ}^{\{1\}}=\{I,X_{i},Y_{i},Z_{i},Z_{i}Z_{j},Z_{i}Y_{j},Y_{i}Y_{j}\},
\end{equation}
\begin{equation}
\mathcal{E}_{YZ}^{\{1\}}=\{I,X_{i},Y_{i},Z_{i},X_{i}X_{j},X_{i}Z_{j},Z_{i}Z_{j}\}.
\end{equation}
These versions of $\mathcal{E}^{\{1\}}$ with the permutations $X\leftrightarrow Z$
and $Y\leftrightarrow Z$ respectively on every qubit. More generally,
we define $\mathcal{E}_{XZ}^{\{t\}}$ and $\mathcal{E}_{YZ}^{\{t\}}$
to be versions of $\mathcal{E}^{\{t\}}$ with the permutations $X\leftrightarrow Z$
and $Y\leftrightarrow Z$ respectively on every qubit. Columns two
and three of Table \ref{tab:AD table} show that exhaustive searches
of $\mathcal{G}_{n}$ using the error sets $\mathcal{E}_{XZ}^{\{1\}}$
and $\mathcal{E}_{YZ}^{\{1\}}$ yield optimal codes for $n=8$ and
$n=9$.

\begin{table}
\caption{\label{tab:AD table} Number of elements of $\mathcal{G}_{n}$ that
yield optimal $((n,K))$ CWS codes for the LC-equivalent error sets
$\mathcal{E}^{\{1\}}$, $\mathcal{E}_{XZ}^{\{1\}}$, and $\mathcal{E}_{YZ}^{\{1\}}$.
The values given for $n=9$ are lower bounds.}
\begin{tabular}{c|ccc}
 & $\mathcal{E}^{\{1\}}$ & $\mathcal{E}_{XZ}^{\{1\}}$ &  $\mathcal{E}_{YZ}^{\{1\}}$\tabularnewline
\hline 
$((5,2))$ & 5 & 9 & 3\tabularnewline
$((6,4))$ & 11 & 16 & 0\tabularnewline
$((7,8))$ & 114 & 157 & 181\tabularnewline
$((8,10))$ & 0 & 4 & 36\tabularnewline
$((9,20))$ & 0 & 6 & 44\tabularnewline
\end{tabular}
\end{table}

\begin{figure}
\includegraphics[scale=0.55]{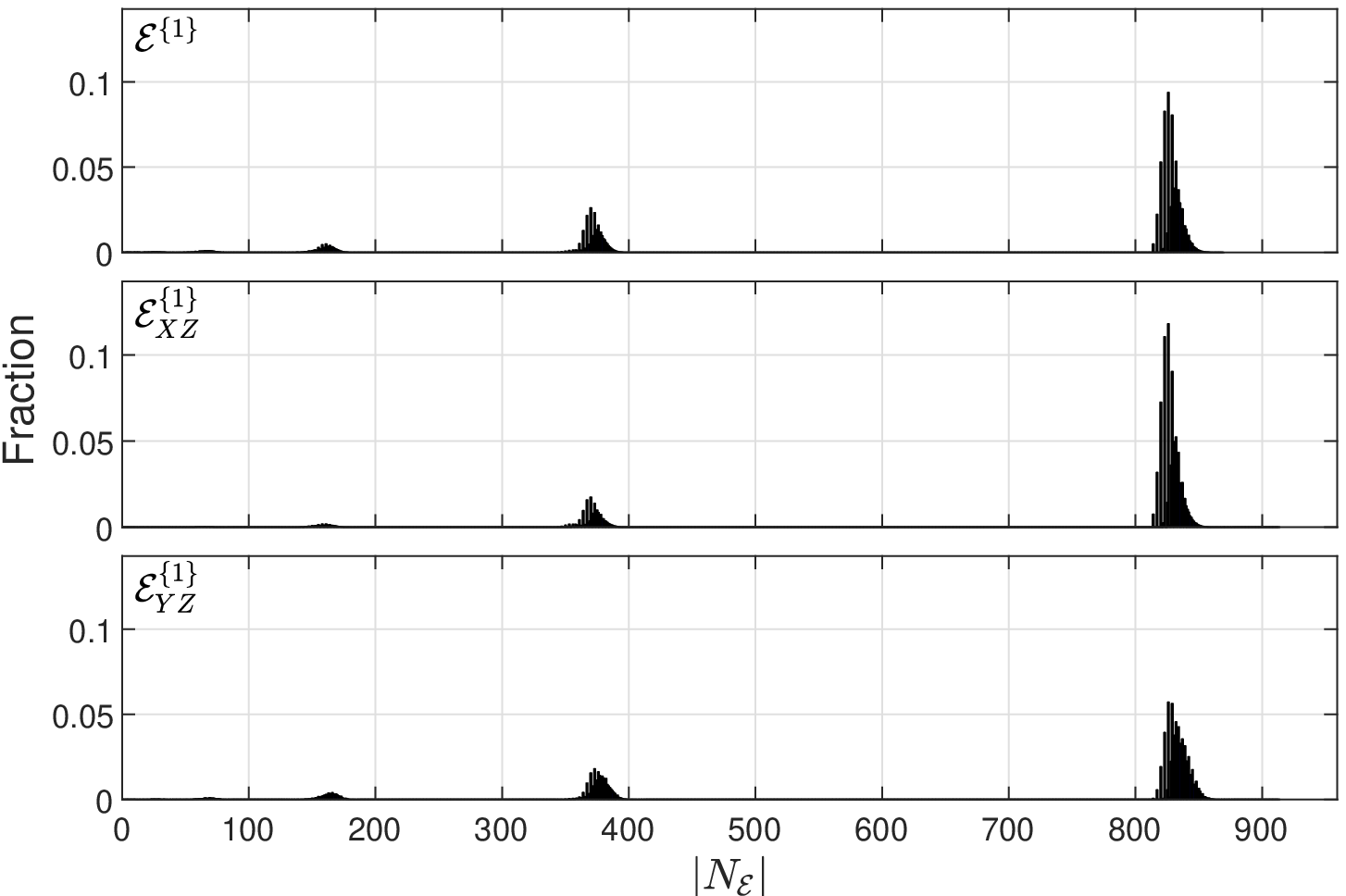}\caption{\label{n10AD1Vhist}Distribution of clique graph order over $\mathcal{G}_{10}$
for the error sets $\mathcal{E}^{\{1\}}$, $\mathcal{E}_{XZ}^{\{1\}}$,
and $\mathcal{E}_{YZ}^{\{1\}}$.}
\end{figure}

For $n=10$, the size of $\mathcal{G}_{10}$ combined with the sizes
of the clique graphs generated makes an exhaustive search impractical.
However, we can still determine the distribution of clique graph sizes
over $\mathcal{G}_{n}$ for the three error sets $\mathcal{E}^{\{1\}}$,
$\mathcal{E}_{XZ}^{\{1\}}$, and $\mathcal{E}_{YZ}^{\{1\}}$ as shown
in Fig. \ref{n10AD1Vhist}. For each of the three error sets, $50,000$
graphs in the $|N_{\mathcal{E}}|>600$ cluster have been selected.
In each case, all $50,000$ graphs yielded $K=32$ codes, which match
the size of the stabilizer code presented in Ref. \citep{gottesman1997stabilizer}.

\begin{figure}
\includegraphics[scale=0.55]{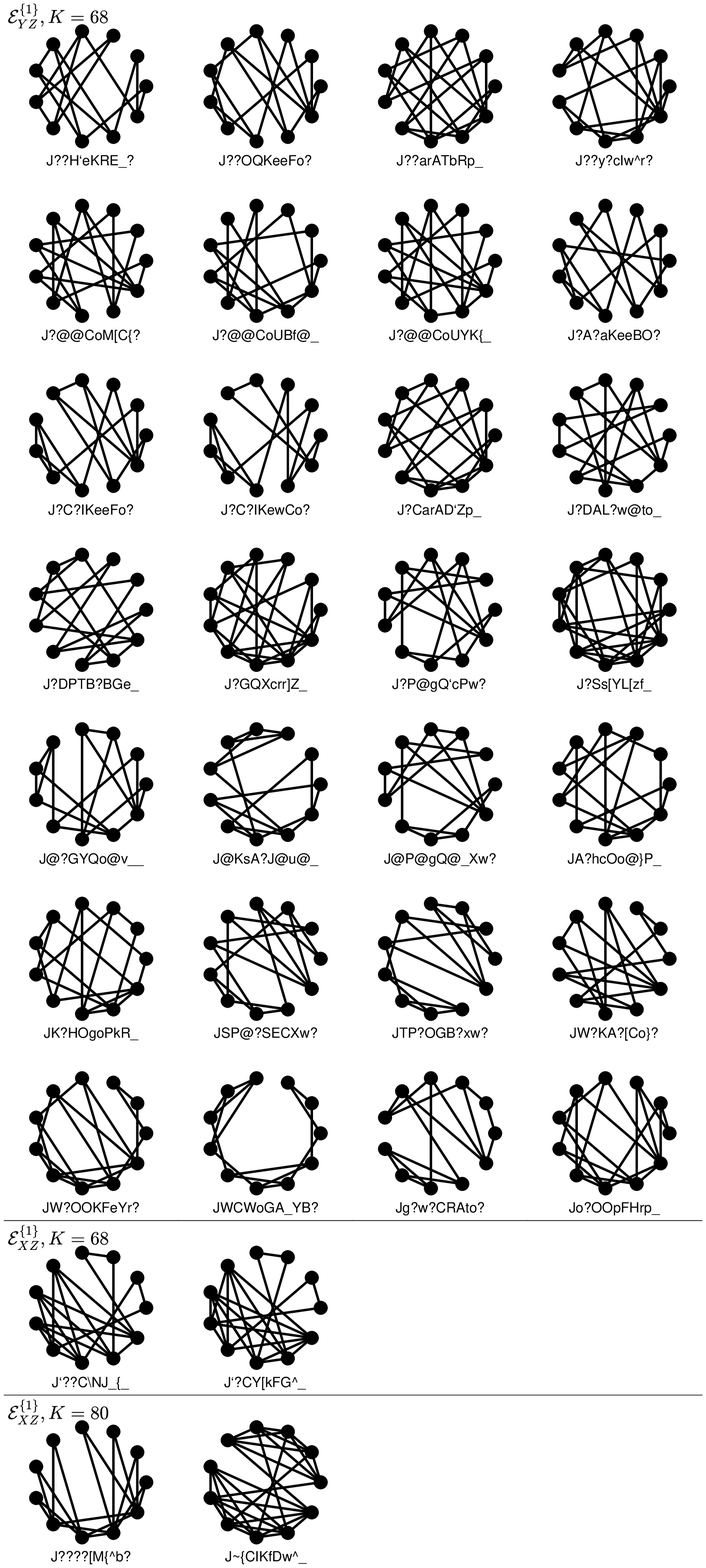}\caption{\label{fig:n11d1p2K68graphs}Nonisomorphic graphs yielding $((11,68))$
codes detecting $\mathcal{E}_{YZ}^{\{1\}}$, $((11,68))$ codes detecting
$\mathcal{E}_{XZ}^{\{1\}}$, and $((11,80))$ codes detecting $\mathcal{E}_{XZ}^{\{1\}}$.}
\end{figure}

For $n=11$, an exhaustive search of $\mathcal{G}_{11}$ is impractical,
even to simply determine clique graph sizes. We have therefore run
$50,000$ instances of our genetic algorithm for each of the three
error sets $\mathcal{E}^{\{1\}}$, $\mathcal{E}_{XZ}^{\{1\}}$, and
$\mathcal{E}_{YZ}^{\{1\}}$. For $\mathcal{E}^{\{1\}}$, this has
yielded a $K=64$ code in every case. These codes match the size of
the stabilizer code presented in Ref. \citep{gottesman1997stabilizer}.
For $\mathcal{E}_{YZ}^{\{1\}}$, $1,818$ instances yielded codes
with $K=68$, which are larger than the best known stabilizer codes.
$28$ of these graphs are nonisomorphic and are shown in Fig. \ref{fig:n11d1p2K68graphs}
(a simple circular node layout is used here as we do not have the
freedom of picking an LC-isomorphic graph that can be drawn clearly
using force-directed layout). For $\mathcal{E}_{XZ}^{\{1\}}$, only
nine instances yielded codes with $K=68$; however, there were also
$71$ instances that yielded codes with $K=80$. Of these, two of
the $K=68$ graphs are nonisomorphic and two of the $K=80$ graphs
are nonisomorphic; these graphs are also shown in Fig. \ref{fig:n11d1p2K68graphs}.
For $n=12$, applying the same genetic algorithm approach has yielded
codes with $K=128$, which match the size of the stabilizer code presented
in Ref. \citep{gottesman1997stabilizer}. In particular, of the $50,000$
instances run for each error set, $21,535$ gave a $K=128$ code for
$\mathcal{E}^{\{1\}}$, $34,906$ gave a $K=128$ code for $\mathcal{E}_{XZ}^{\{1\}}$,
and $41,002$ gave a $K=128$ code for $\mathcal{E}_{YZ}^{\{1\}}$.

\subsection{Two amplitude damping errors}

As determined by exhaustive search in Ref. \citep{jackson2016codeword},
there are no nontrivial CWS codes capable of detecting the error set
$\mathcal{E}^{\{2\}}$ with $n\leq8$. For $n=9$, the largest CWS
code that can detect $\mathcal{E}^{\{2\}}$ has $K=2$. Interestingly,
an exhaustive search of $\mathcal{G}_{9}$ fails to yield any $K=2$
codes detecting $\mathcal{E}^{\{2\}}$. However, there are seven elements
of $\mathcal{G}_{9}$ that yield $K=2$ codes detecting $\mathcal{E}_{XZ}^{\{2\}}$
and $12$ elements that yield $K=2$ codes detecting $\mathcal{E}_{YZ}^{\{2\}}$.
For $n=10$, there are $32$ elements of $\mathcal{G}_{10}$ that
yield a $K=2$ code detecting $\mathcal{E}^{\{2\}}$, $309$ that
yield a $K=2$ code detecting $\mathcal{E}_{XZ}^{\{2\}}$, and $1,327$
that yield a $K=2$ code detecting $\mathcal{E}_{YZ}^{\{2\}}$. There
are no larger standard form $n=10$ codes detecting $\mathcal{E}^{\{2\}}$,
$\mathcal{E}_{XZ}^{\{2\}}$, or $\mathcal{E}_{YZ}^{\{2\}}$. 

\begin{figure}
\includegraphics[scale=0.55]{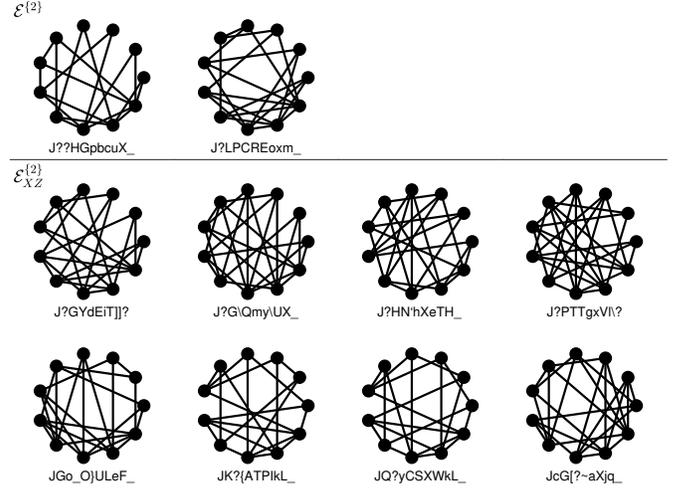}\caption{\label{fig:n11d2ADgraphs}Nonisomorphic graphs yielding $((11,4))$
codes detecting either $\mathcal{E}^{\{2\}}$ or $\mathcal{E}_{XZ}^{\{2\}}$.}
\end{figure}

\begin{figure}
\includegraphics[scale=0.55]{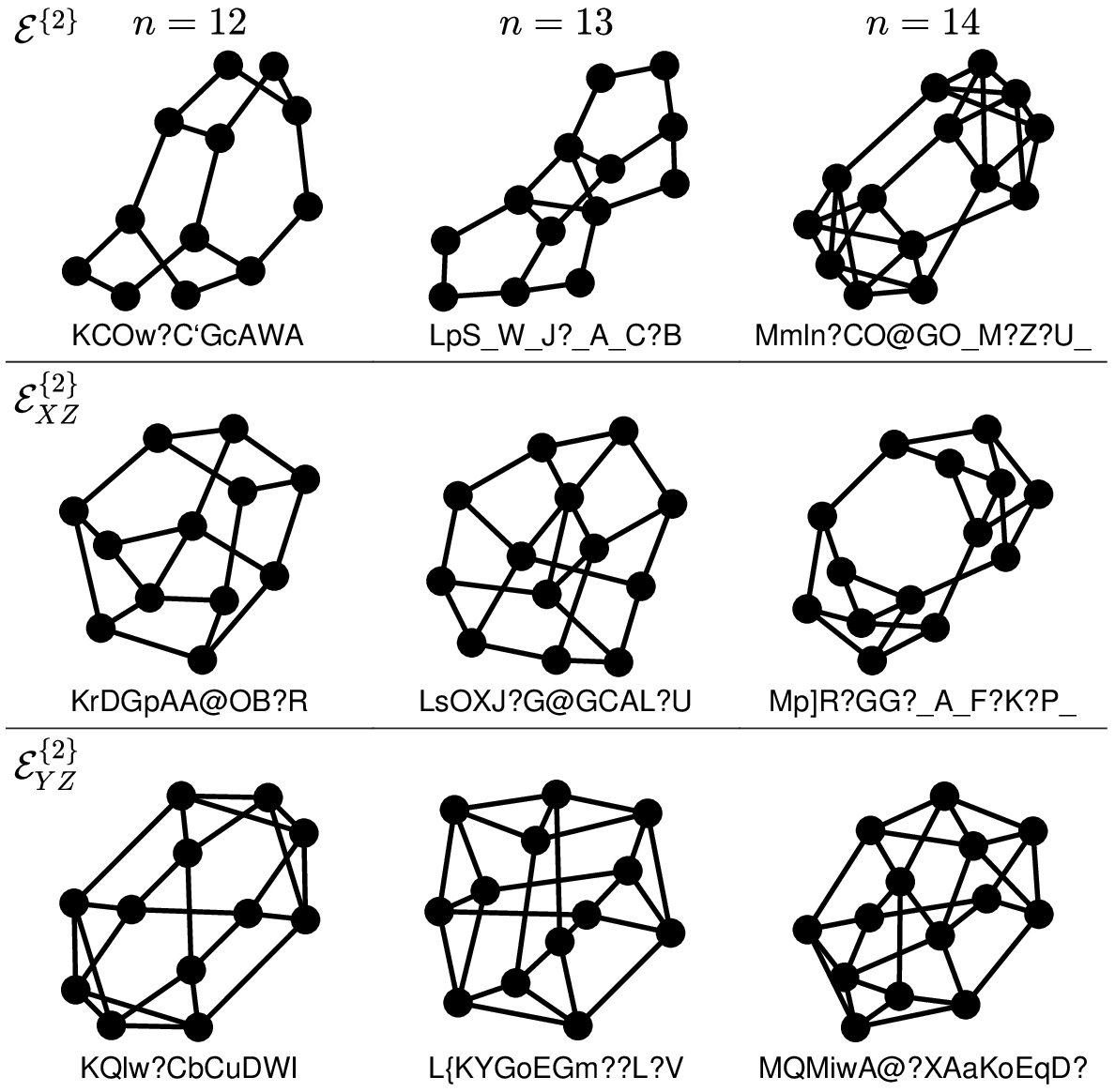}\caption{\label{fig:n12-14ADgraphs}Graphs yielding $((12,4))$, $((13,8))$,
or $((14,16))$ codes detecting one of $\mathcal{E}^{\{2\}}$, $\mathcal{E}_{YZ}^{\{2\}}$,
or $\mathcal{E}_{XZ}^{\{2\}}$. }
\end{figure}

As in the single error correcting case, any exhaustive search of $\mathcal{G}_{n}$
for $n\geq11$ is impractical. For $11\leq n\leq14$, we have run
$50,000$ instances of the genetic algorithm outlined in Sec. \ref{subsec:Distance-four-codes}
for each of the three error sets $\mathcal{E}^{\{2\}}$, $\mathcal{E}_{XZ}^{\{2\}}$,
and $\mathcal{E}_{YZ}^{\{2\}}$. The best codes found have $K=4$
for $n=11$ and $n=12$, $K=8$ for $n=13$, and $K=16$ for $n=14$.
The number of genetic algorithm instances yielding codes with these
parameters is shown in Table \ref{tab:two AD table}. Note that nearly
all of these codes are stabilizer codes. For $n\leq13$, we have used
an exact clique finder, whereas for $n=14$ we have used PLS. The
$n=11$ codes are interesting due to how difficult they are to find.
The two graphs found for $\mathcal{E}^{\{2\}}$ are nonisomorphic
and eight of those found for $\mathcal{E}_{XZ}^{\{2\}}$ are nonisomorphic.
These graphs are shown in Fig. \ref{fig:n11d2ADgraphs}. It is easy
to find graphs giving codes with $K=4$ codes for $n=12$, $K=8$
for $n=13$, and $K=16$ for $n=14$ (they can be found quickly even
with a simple random search). However, to the best of our knowledge
no stabilizer codes with these parameters have been previously published.
Furthermore, they are all larger than an optimal $d=5$ stabilizer
codes that can correct two arbitrary errors. As such, we include graphs
yielding codes of these sizes for $\mathcal{E}^{\{2\}}$, $\mathcal{E}_{XZ}^{\{2\}}$,
and $\mathcal{E}_{YZ}^{\{2\}}$ in Fig. \ref{fig:n12-14ADgraphs}.

\begin{table}[b]
\caption{\label{tab:two AD table} The number of genetic algorithm instances
out of the $50,000$ run that yielded an $((n,K))$ code detecting
the given error set.}
\begin{tabular}{c|ccc}
 & $\mathcal{E}^{\{2\}}$ & $\mathcal{E}_{XZ}^{\{2\}}$ &  $\mathcal{E}_{YZ}^{\{2\}}$\tabularnewline
\hline 
$((11,4))$ & $2$ & $14$ & $0$\tabularnewline
$((12,4))$ & $45,912$ & $36,275$ & $43,225$\tabularnewline
$((13,8))$ & $38,475$ & $33,163$ & $44,151$\tabularnewline
$((14,16))$ & $3,467$ & $5,840$ & $13,148$\tabularnewline
\end{tabular}
\end{table}

\section{Conclusion\label{sec:Conclusion}}

We have demonstrated the effectiveness of a number of heuristic approaches
to the construction of CWS codes. We have shown that using an approximate
maximum clique finding algorithm makes finding larger codes practical.
In particular, this has allowed us to find $((9,97\leq K\leq100,2))$
and $((11,387\leq K\leq416,2))$ codes that are larger than the best
known nonadditive codes. We have demonstrated a clustering of clique
graph sizes and shown a relationship between clique graph order and
code size. Furthermore, we have shown that graphs yielding large clique
graphs can be found using a genetic algorithm with a novel spectral
bisection based crossover operation. This search strategy has yielded
$((13,18,4))$ and $((13,20,4))$ codes, which are larger than any
previously known code. Finally, we have shown that good asymmetric
codes can be found by considering standard form codes that detect
one of only three of the $3^{n}$ possible LC-equivalent error sets.
Coupling this with the genetic algorithm approach, we have found $((11,68))$
and $((11,80))$ codes capable of correcting a single amplitude damping
error. We have also found $((11,4))$, $((12,4))$, $((13,8))$, and
$((14,16))$ stabilizer codes capable of correcting two amplitude
damping errors.

\bibliographystyle{apsrev4-1}
\bibliography{cwsPaper}
\FloatBarrier
\end{document}